\documentclass[conference]{IEEEtran}

\usepackage{algorithm}
\usepackage{algpseudocode}
\usepackage{amsmath}
\usepackage{amssymb}
\usepackage{mathrsfs}
\usepackage{multirow}
\usepackage{threeparttable}
\usepackage{tabularx}
\usepackage{colortbl}
\usepackage{booktabs}

\usepackage{enumitem}
\usepackage{pifont}
\usepackage{subfigure}
\usepackage{graphicx}
\usepackage{xcolor}
\usepackage{setspace}
\usepackage{xspace}
\usepackage{hyperref}
\usepackage{marvosym}

\usepackage{layout}

\definecolor{deepblue}{RGB}{54,79,106}
\definecolor{orange}{RGB}{203, 124, 89}
\definecolor{teclisblue}{RGB}{160,181,226}
\definecolor{red}{RGB}{177,36,25}
\definecolor{fullblue}{RGB}{72,120,208}

\newcommand{\red}[1]{\textcolor{red}{#1}}
\newcommand{\blue}[1]{\textcolor{fullblue}{#1}}

\newcommand{\this}{NetMasquerade\xspace}
\newcommand{\model}{Traffic-BERT\xspace}
\newcommand{\etal}{{\textit{et al. }}}
\newcommand{\tabincell}[2]{\begin{tabular}{@{}#1@{}}#2\end{tabular}}

\begin{document}
\title{A Hard-Label Black-Box Evasion Attack against ML-based Malicious Traffic Detection Systems}

\author{
\IEEEauthorblockN{
    Zixuan Liu\IEEEauthorrefmark{1},
    Yi Zhao\IEEEauthorrefmark{2}\textsuperscript{\Letter},
    Zhuotao Liu\IEEEauthorrefmark{1}\IEEEauthorrefmark{3},
    Qi Li\IEEEauthorrefmark{1}\IEEEauthorrefmark{3}, 
    Chuanpu Fu\IEEEauthorrefmark{1}, 
    Guangmeng Zhou\IEEEauthorrefmark{1}, 
    Ke Xu\IEEEauthorrefmark{1}\IEEEauthorrefmark{3}\textsuperscript{\Letter}}

\IEEEauthorblockA{
    \IEEEauthorrefmark{1}Tsinghua University,
    \IEEEauthorrefmark{2}Beijing Institute of Technology,
    \IEEEauthorrefmark{3}Zhongguancun Lab
}
\vspace{-8cm}
}

\IEEEoverridecommandlockouts
\makeatletter\def\@IEEEpubidpullup{4.0\baselineskip}\makeatother
\IEEEpubid{\parbox{\columnwidth}{
		Network and Distributed System Security (NDSS) Symposium 2026\\
		23 - 27 February 2026 , San Diego, CA, USA\\
		ISBN 979-8-9919276-8-0\\  
		https://dx.doi.org/10.14722/ndss.2026.240916\\
		www.ndss-symposium.org
}
\hspace{\columnsep}\makebox[\columnwidth]{}}

\maketitle

\begin{abstract}

Machine Learning (ML)-based malicious traffic detection is a promising security paradigm. It outperforms rule-based traditional detection by identifying various advanced attacks.
However, the robustness of these ML models is largely unexplored, thereby allowing attackers to craft adversarial traffic examples that evade detection.
Existing evasion attacks typically rely on overly restrictive conditions (e.g., encrypted protocols, Tor, or specialized setups), or require detailed prior knowledge of the target (e.g., training data and model parameters), which is impractical in realistic black-box scenarios. The feasibility of a hard-label black-box evasion attack (i.e., applicable across diverse tasks and protocols without internal target insights) thus remains an open challenge.

To this end, we develop \this, which leverages reinforcement learning (RL) to manipulate attack flows to mimic benign traffic and evade detection. Specifically, we establish a tailored pre-trained model called
\model, utilizing a network-specialized tokenizer and an attention mechanism to extract diverse benign traffic patterns. Subsequently, we integrate \model into the RL framework, allowing \this to effectively manipulate malicious packet sequences based on benign traffic patterns with minimal modifications. 
Experimental results demonstrate that \this enables both brute-force and stealthy attacks to evade 6 existing detection methods under 80 attack scenarios, achieving over 96.65\% attack success rate. Notably, it can evade the methods that are either empirically or certifiably robust against existing evasion attacks. Finally, \this achieves low-latency adversarial traffic generation, demonstrating its practicality in real-world scenarios.

\end{abstract}

\section{Introduction} \label{chapter:introduction}

Machine learning (ML)-based malicious traffic detection systems identify attack behaviors by learning the features of traffic~\cite{CCS21-Whisper, NDSS18-Kitsune, NDSS23-HV}. As an emerging security paradigm, it is promising for identifying
multiple sophisticated attacks~\cite{S&P13-crossfire, SIGCOMM03-low-TCP, CCS13-distributedSSH} and thus outperforms traditional rule-based detection~\cite{SEC23-xNIDS, Security01-NIDSSurvey, CCS23-Alert-Alchemy} in both effectiveness~\cite{CCS22-TorDetection, AISEC16-Cisco} and efficiency~\cite{Security23-NetBeacon, NDSS21-FlowLens}. Currently,
ML-based systems are deployed to complement the traditional systems due to their ability to detect unknown~\cite{CCS17-Deeplog} and encrypted~\cite{AISEC16-Cisco, CONEXT12-BotFinder} attack traffic. 

Unfortunately, as in many other ML-based domains~\cite{ICLR14-FGSM,CVPR15-Nguyen,ICLR15-FGSM}, robustness issues are prevalent in ML-based traffic detection systems~\cite{SP10-Outside,SEC22-Dos-Donots,CCS22-Emperor}. 
That is, attackers can craft adversarial traffic by adding, delaying, or otherwise modifying packets~\cite{JSAC21-TM, TDSC22-Manda}, causing detection models to misclassify these deceptive flows as benign. 
The research community has put forward a range of advanced evasion methods (see Table~\ref{tab: method_comparison}), yet many of these methods operate under narrowly defined conditions, such as leveraging encrypted protocols~\cite{Security21-BLANKET, CoNEXT23-Amoeba, CCS21-DomainConstraintsRobustness}, tunnels~\cite{Security21-BLANKET}, Tor networks~\cite{TIFS24-rudolf}, or circumventing specialized third-party censorship systems~\cite{CoNEXT23-Amoeba}. The effectiveness of these protocol-related and task-related approaches drops significantly when the attack environment changes. Moreover, some solutions rely heavily on prior knowledge about the target or dataset, requiring full (white-box)~\cite{CCS21-DomainConstraintsRobustness, Security21-BLANKET, TDSC22-Manda} or partial (gray-box)~\cite{Security21-BLANKET, JSAC21-TM} access, which is impractical in a more realistic black-box setting.

To bridge these gaps, we aim to design a black-box adversarial attack targeting widely used ML-based traffic detection systems that rely on statistical patterns~\cite{CCS21-Whisper, NDSS18-Kitsune, Security23-NetBeacon, NDSS21-FlowLens}. In particular, the attack must be protocol-agnostic and task-agnostic, allowing it to be seamlessly applied to any malicious traffic, regardless of whether it is encrypted, tunneled, or otherwise constrained. Moreover, the attacker can generate adversarial malicious traffic with minimal modifications, relying solely on whether the target system drops malicious packets (i.e., hard-label attack~\cite{S&P23-BounceAttack, Security23-PatchAttack}). In contrast to feature-space attacks~\cite{PAKDD22-Idsgan, ESA21-Alhajjar}, i.e., impractical settings that require attackers to interfere with ML execution, our traffic modifications must preserve the effectiveness of the attacks~\cite{CCS21-DomainConstraintsRobustness, JSAC21-TM}.

\newcommand{\yesmark}{\ding{51}}
\newcommand{\nomark}{\ding{55}}

\newcommand{\y}{\blue{\yesmark}}
\newcommand{\n}{\red{\nomark}}

\newcommand{\req}{\red{Required}}
\newcommand{\nreq}{\blue{Not Req.}}

\renewcommand{\arraystretch}{1.25}
\begin{table*}[t]
    \small
    \centering
    \caption{\textbf{Comparison of Existing Evasion Attacks against Traffic Analysis Systems.}}
    \vspace{-2mm}
    \resizebox{\textwidth}{!}{
    \begin{threeparttable}
    \begin{tabular}{@{}c|c|cc|ccc|cc@{}}
        \toprule
        \multirow{2}{*}{Scenarios} &    \multirow{2}{*}{Evasion Techniques}            & \multicolumn{2}{c|}{Attack Applicability} & \multicolumn{3}{c|}{Without Prior Knowledge} & \multicolumn{2}{c}{Attack Performance} \\
        \cline{3-9}
                                    &                                                    & Protocol-agnostic & Task-agnostic   & Datasets & Features & Model          & Low Overhead & Low Latency \\
        \midrule
        \multirow{2}{*}{White-box} & Gradient Analysis~\cite{CCS21-DomainConstraintsRobustness, Security21-BLANKET}   & \n & \y                         & \n   & \n   & \n               & \n & \y \\
        & Optimization~\cite{TDSC22-Manda}                                                 & \n & \y                         & \n   & \n   & \n                   & \n & \y \\
        \midrule
        \multirow{2}{*}{Gray-box} & Sample Transferability~\cite{Security21-BLANKET}       & \n & \n                         & \n   & \n   & \y                  & \n & \y \\
        & Feature Manipulation~\cite{JSAC21-TM}                                            & \y & \y                         & \y   & \n   & \y                 & \n & \n \\
        \midrule
        \multirow{2}{*}{Black-box} & Packet reassembly~\cite{CoNEXT23-Amoeba}              & \n & \n                         & \y   & \y   & \y                & \n & \n \\
        & Traffic Mimicking \textbf{(Ours)}                                                & \y & \y                         & \y   & \y   & \y                & \y & \y \\
        \bottomrule
    \end{tabular}
    \end{threeparttable}
    \label{tab: method_comparison}
    }
\end{table*}

This paper presents \this, a hard-label black-box evasion attack, which utilizes deep reinforcement learning (RL) to transform malicious traffic into adversarial examples by mimicking benign traffic patterns. At its core, we propose \model, a tailored pre-trained model for capturing diverse and complex benign traffic distributions. Subsequently, we develop an RL framework that decides the location and type of packet modification step-by-step, leveraging \model's embedded knowledge of benign behaviors. The only feedback required for the RL training process is the blocked-or-not signal from the targeted detection system. The detailed design ensures that \this achieves minimal, yet effective, modifications across diverse traffic types and detection models, thereby evading detection systems under black-box conditions. We address two main challenges in constructing effective adversarial traffic.

First, we must capture rich benign traffic patterns in order to mimic them convincingly. To address this, we study the distribution of Internet packet patterns, then pad and chunk traffic from public datasets using an optimal setup to improve diversification. Afterwards, we pre-process the traffic with network-specific tokenizers. Finally, we extract dependencies among the tokens with a novel attention block in \model, providing a robust representation of benign traffic across various protocols and scenarios.

Second, the RL model for generating optimal evasion policies must maintain both low training overhead and low online inference latency. To this end, we formulate the traffic modifications as a Finite Horizon Markov Decision Process, enabling a multi-step decision strategy with explicit incentives for minimal and targeted modifications, effectively reducing inference costs. Meanwhile, we utilize a lightweight policy network and ensure rapid convergence by leveraging already-learned benign distributions in \model, significantly reducing training overhead. In addition, we introduce an effectiveness penalty, which safeguards the malicious functionality of the attack.

We prototype \this\footnote{Source code: https://github.com/09nat/NetMasquerade} with Pytorch~\cite{Pytorch} and Intel DPDK~\cite{DPDK}. Experiments demonstrate that \this enables both high-rate and low-rate attacks to evade $6$ top-performing detection systems in $80$ attack scenarios, achieving over $96.65\%$ attack success rate (ASR). Note that \this can evade the methods that are either empirically~\cite{CCS21-Whisper} or certifiably~\cite{NDSS23-BARS} robust against existing evasion attacks. 
Compared with other attacks~\cite{CCS21-Whisper, JSAC21-TM}, \this applies minimal modifications to no more than $10$ steps in all test scenarios, thus incurring little impact, e.g., a Kullback-Leibler (KL) divergence of $0.013$ between the original and modified bandwidth distributions.
Moreover, the evasion attacks can be constructed in time, i.e., \this can transform 4.239K adversarial packets per second. 

In general, our studies reveal that the robustness issue of traffic detection remains unaddressed, which emphasizes the necessity of enhancing robustness against advanced evasion attacks. The contributions of this paper are three-fold:

\begin{itemize}
    \item We develop a hard-label black-box evasion attack that operates across diverse traffic types and targets widely used ML-based detection systems relying on statistical patterns, leveraging deep reinforcement learning to manipulate attack packets for efficient benign-traffic mimicry.
    \item  We design a tailored pre-trained model, \model, to capture diverse and complex benign traffic patterns, equipping NetMasquerade with the ability to transform malicious flows into benign-like traffic across a wide range of protocols and tasks.
    \item We establish an RL-based framework that transforms original attack traffic into adversarial examples with minimal modifications, and experimentally validate that \this can generate various attack traffic to evade multiple state-of-the-art detection systems with small overhead.
\end{itemize}
\begin{figure}[t]   
    \centering
    \includegraphics[width=\columnwidth]{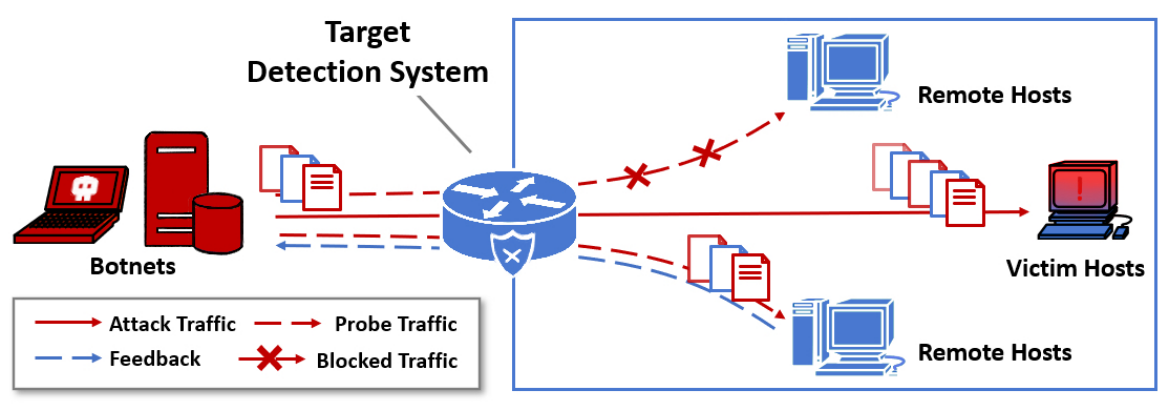}
    \vspace{-4mm}
    \caption{\textbf{Network topology.}}
    \vspace{-4mm}
    \label{fig:topo}
  \end{figure}

\begin{figure*}[t]
    \centering
    \includegraphics[scale=0.48]{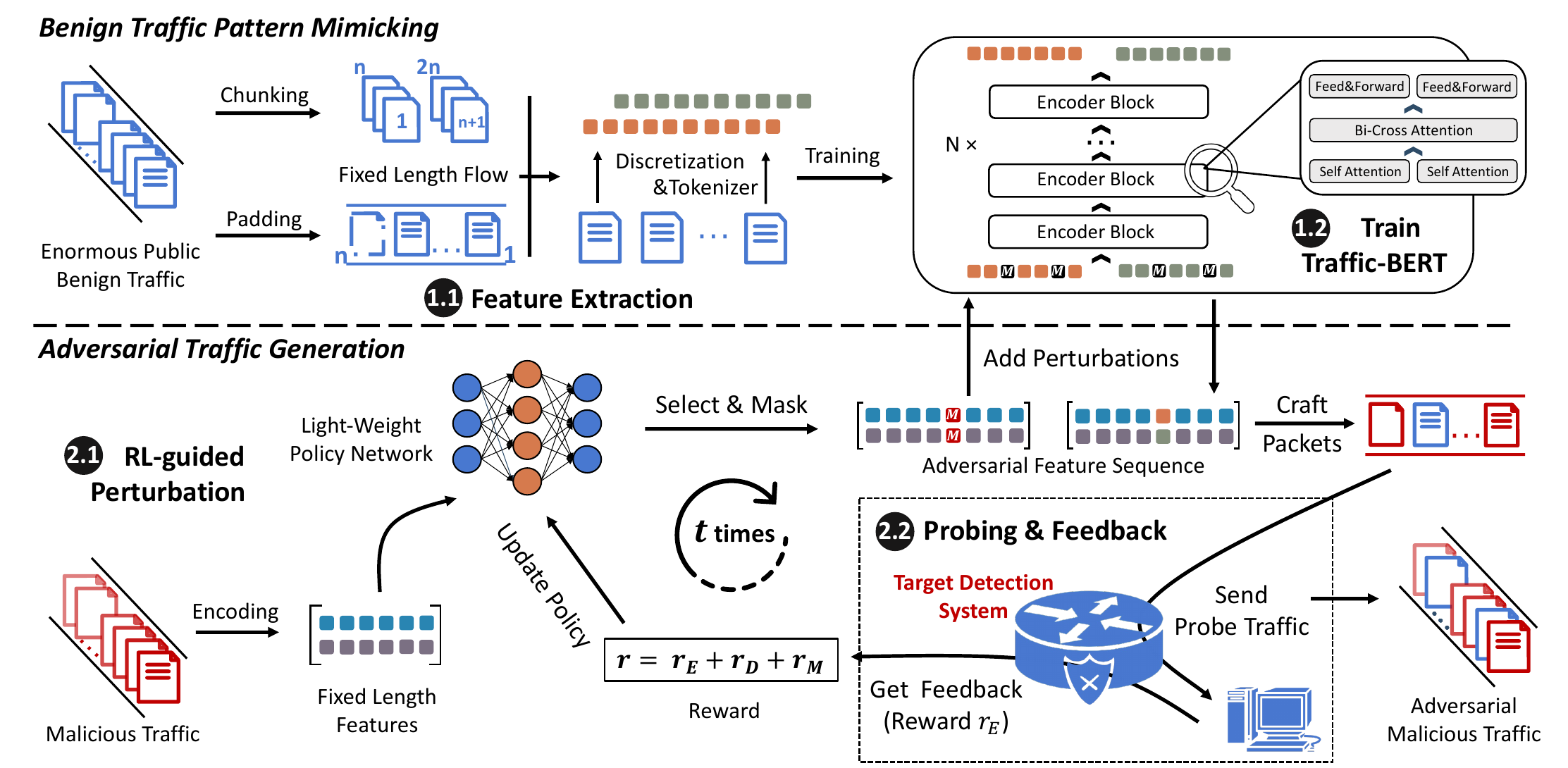}
    \vspace{-6mm}
    \caption{\textbf{High-Level Design of \this.}}
    \vspace{-3mm}
    \label{fig:High-Level Design}
    \vspace{-1mm}
\end{figure*}

\section{Threat Model and Assumptions} \label{chapter:threat}

\noindent\textbf{Threat Model.} We assume an external adversary who instructs botnets to deliver malicious traffic (e.g., high-speed attack and stealthy malware behaviors) toward victim hosts~\cite{CONEXT12-BotFinder}. The attack flows originate on the public Internet and must traverse an in-line, ML-based traffic detection system deployed at the victim's ingress link, as shown in Figure~\ref{fig:topo}. 
The detection system operates on certain flow‑ or packet‑level statistical patterns (e.g., sizes, delays) for traffic classification and does not inspect the payload~\cite{CCS21-Whisper, NDSS23-HV, NDSS18-Kitsune, NDSS21-FlowLens}. Such pattern-based models are increasingly popular as they are effective on encrypted traffic. Meanwhile, the system forwards benign traffic and drops or rate-limits malicious traffic. This behavior is consistent with both  existing academic proposals~\cite{NDSS20-Poseidon} and default configurations of open-source in-line detectors~\cite{suricata}, which prioritize network availability by choosing low-latency actions like packet dropping over more aggressive responses (e.g., blocking entire source IPs, which could be a NAT gateway serving many legitimate users) that may cause significant collateral damage. To evade detection, the attacker craft each malicious flow to construct adversarial traffic that misleads detection systems into classifying them as benign, while retaining the flow's original intent.

\noindent\textbf{Assumptions.} The attacker cannot obtain any details of the target detection systems, such as ML models, parameters, feature extractors and training datasets. That means, the attacker treats target systems as a strict (i.e., hard-label) black-box,  which differs from traditional white-box and gray-box attacks that either need to access ML models (e.g., obtain gradients)~\cite{TDSC22-Manda} or datasets~\cite{Security21-BLANKET} for better transferability. This black-box setting meets the real-world scenarios, as most traffic detection systems are closed-source software~\cite{Cisco2021encrypted} or outsourced cloud services~\cite{Cloudflare-DDoS-Products,AKamai-Prolexic}, effectively preventing attackers from obtaining any information.

However, the attacker can conduct a reconnaissance phase to gather pass/fail feedback from the target detection system. Specifically, the attacker sends probe traffic to remote hosts behind the target detection systems. The probe traffic should exactly mirror the malicious traffic's intended traffic pattern (i.e., matching packet sizes and inter-packet delays) without embedding the original malicious payload (i.e., the attacker can freely embed payloads of the same size). For TCP flows, any return packet (e.g., RST, ACK, SYN-ACK) from the destination~\cite{NDSS09-Weaver} signals that the probe has traversed the IDS, whereas the complete absence of a reply within the retransmission window indicates the blockage~\cite{RFC9293-TCP}. 
For UDP flows, the attacker could employ a stateful application-layer protocol (e.g., QUIC~\cite{RFC9000-QUIC}) to induce a response. If the destination port is closed, the attacker would typically observe ICMP Unreachable when the traffic successfully passes the detection~\cite{RFC0792-ICMP,Nmap}. However, no such message will be received if the flow is blocked. 
By checking for a response within a fixed timeout, the attacker assigns a pass/fail label to each probe. 
Besides, side-channel techniques (e.g., hop-limited TTL~\cite{IMC13-Tracebox,Traceroute}, IPID~\cite{CCS20-IPIDsidechannel,SP12-TCPseqinfer}) can further reveal whether the traffic reaches the destination. Overall, this binary indicator (i.e., hard-label) enables the attacker to refine subsequent adversarial traffic generation. In addition, the attacker can access benign traffic from public datasets~\cite{WIDE2023}, which differ from those used to train the traffic detection model. We supplement additional discussions regarding our assumptions in \S~\ref{chapter:evaluation}.
\section{The Overview of \this} \label{chapter:overview}

\subsection{Key Observation}
We begin by noting that modern traffic detection~\cite{AKamai-Prolexic,Cisco2021encrypted,Cloudflare-DDoS-Products} operates as a strict (i.e., hard-label) black-box system from the perspective of typical malicious attackers. All model implementations and training details are concealed, which significantly degrades the performance of existing white-box~\cite{Security21-BLANKET,CCS21-DomainConstraintsRobustness} and gray-box~\cite{JSAC21-TM,Security21-BLANKET} methods. In contrast, such systems often drop or throttle~\cite{NDSS20-Poseidon,SEC21-Ripple,CCS24-Exosphere} malicious traffic while allowing benign traffic to pass. This behavioral asymmetry inherently yields a feedback signal, motivating us to interactively probe the detector's decision boundary and steer malicious traffic within that boundary. We accomplish this process with reinforcement learning.

However, Internet traffic spans a wide variety of protocols and use cases. A naive design can degrade RL performance (see \S~\ref{subsection: Deep Dive Appendix}), while task-specific attacks fail in heterogeneous environments~\cite{CoNEXT23-Amoeba}. Fortunately, existing studies show that benign traffic distributions tend to be denser, whereas malicious traffic is often more sparse~\cite{NDSS24-RAPIER}. This density gap encourages us to morph the malicious flow so that its features migrate toward the benign manifold while preserving attack semantics. To achieve this, we develop an effective semantic model (i.e., Traffic-BERT) to capture benign traffic patterns, thereby guiding RL training through its learned representations. Finally, we introduce a two-stage divide-and-conquer training framework, along with several additional mechanisms, to significantly reduce the overhead introduced by integrating Traffic-BERT with the RL process while preserving the effectiveness of the generated adversarial traffic.

\subsection{High-Level Architecture}

Figure~\ref{fig:High-Level Design} shows the two stages of \this, a black-box evasion attack method. In the first stage, it captures the benign traffic patterns. In the second stage, it generates adversarial traffic based on the traffic patterns.

\noindent \textbf{Benign Traffic Pattern Mimicking.} In this stage, we focus on comprehensively modeling benign traffic distributions to provide a solid foundation for subsequent adversarial flow generation. To this end, we propose \model, a variant of BERT~\cite{NAACL-HLT18-BERT}, capable of processing multiple feature sequence inputs and outputs. Specifically, we first extract basic flow features (i.e., packet size sequence and inter-packet delay sequence). Next, we introduce dedicated feature extraction and embedding schemes to reconcile the gap between continuous, variable-length traffic data and the fixed-length, discrete input format typically required by \model. Building on these enriched representations, we propose a cross-feature bidirectional attention mechanism to simultaneously capture global dependencies within each individual feature sequence and across heterogeneous feature modalities. By training \model under a Mask-Fill task, we enable it to learn deep bidirectional dependencies and acquire the capability to contextually complete fine-grained benign features. The trained \model can be directly used in \textbf{Adversarial Traffic Generation} to guide the RL optimization process. We will detail the Benign Traffic Pattern Mimicking in \S~\ref{behavior cloning}.

\noindent \textbf{Adversarial Traffic Generation.} In this stage, our goal is to embed the pattern of benign traffic into malicious traffic with minimal modifications while preserving attack semantics and domain constraints. We model this as a standard Markov Decision Process (MDP), and employ deep RL to address complex sequential decision-making. 
Specifically, \this utilizes the Gated Recurrent Units (GRUs)~\cite{NIPS14workshop-GRU}, a lightweight neural network, as the policy network and the state-value networks (a.k.a. Q-Networks). This design significantly reduces training time and inference latency while still effectively capturing temporal flow features. By learning an optimal policy to select packet-level feature positions for masking, \this leverages \model to fill the masked tokens with benign traffic patterns. The resulting adversarial flow is used to probe the target system. The response provides the core feedback, which we integrate with two novel penalty terms to form a comprehensive reward signal: a \textit{dissimilarity penalty}, which ensures that the final adversarial flow remains close to the original malicious flow while also reducing the required inference steps, and an \textit{effectiveness penalty}, which retains the underlying attack function. This complete reward signal then guides the optimization of the policy network using the Soft Actor-Critic (SAC)~\cite{ICML18-SAC} algorithm. We will detail the Adversarial Traffic Generation in \S~\ref{traffic generation}.

\noindent \textbf{Two-Stage Framework Advantages.}  In many RL applications, models are initialized from expert demonstrations via behavior learning and then deployed as policy networks in downstream tasks~\cite{AAAI18-RLfD, Arxiv17-DDPGfD}. However, this \textit{Pretraining-Finetuning} framework is not suitable for the traffic domain because it introduces significant overhead. By contrast, our design cleanly decouples benign traffic modeling (Stage $1$) from adversarial RL optimization (Stage $2$). \model learns high-fidelity benign traffic embeddings without entangling with the RL process, avoiding repeated large-scale retraining. Meanwhile, the lightweight policy network incrementally references the embeddings to weave benign patterns into malicious flows, preserving both the efficiency and the effectiveness of the generated adversarial traffic. 
\section{Benign Traffic Pattern Mimicking}\label{chapter:stage1}
\label{behavior cloning}

\subsection{Feature Extraction} \label{chapter: extraction}
\label{feature extraction}

The feature extraction module encodes network traffic into \model's token sequences. Although various traffic studies have explored related encoding strategies in other contexts~\cite{CCS24-Exosphere,WWW22-ET-BERT,NDSS24-RAPIER,SP24-Trafficformer}, they are not directly applicable to \model for two reasons. First, as a language model, \model demands fixed-length inputs, while real-world flows vary widely in size and duration. It is essential to set a base input length that accommodates the majority of flows and provide a mechanism to capture extended flows without information loss. Second, \model requires tokens to reside in a uniformly discrete space, whereas raw network features (e.g., inter-packet delays) may be highly skewed or continuous. We overcome these issues and characterize flows based on statistical insights, as detailed below.

\begin{figure}[t]
    \subfigcapskip=-1mm
    \vspace{-4mm}
    \begin{center}
    \subfigure [Flow length distribution.]{
        \label{fig:flow length distribution}
		\includegraphics[width=0.26\textwidth]{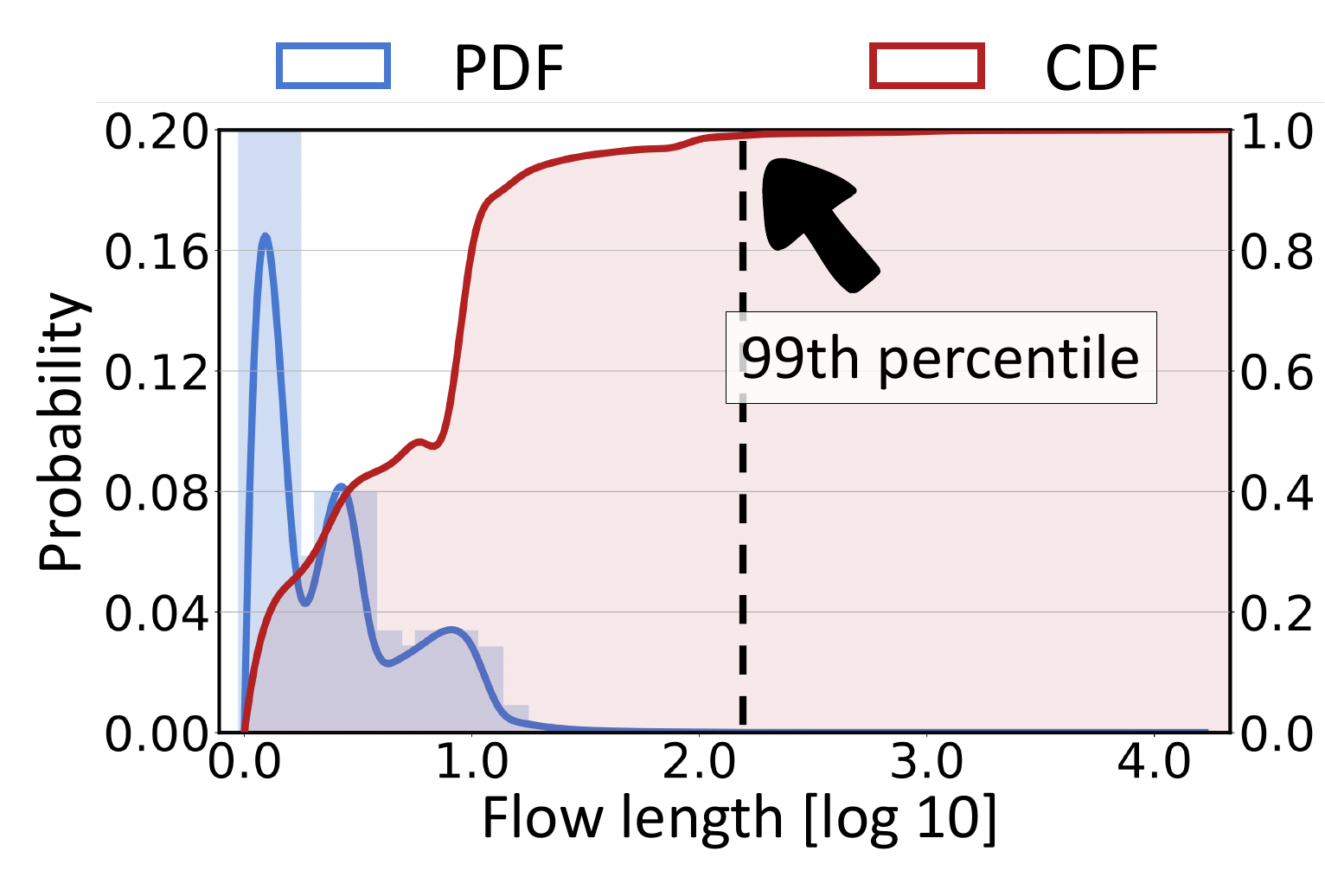}
	}
    \hspace{-3mm}
    \subfigure [Flow entropy ratios.]{
        \label{fig:entropy distribution}
		\includegraphics[width=0.19\textwidth]{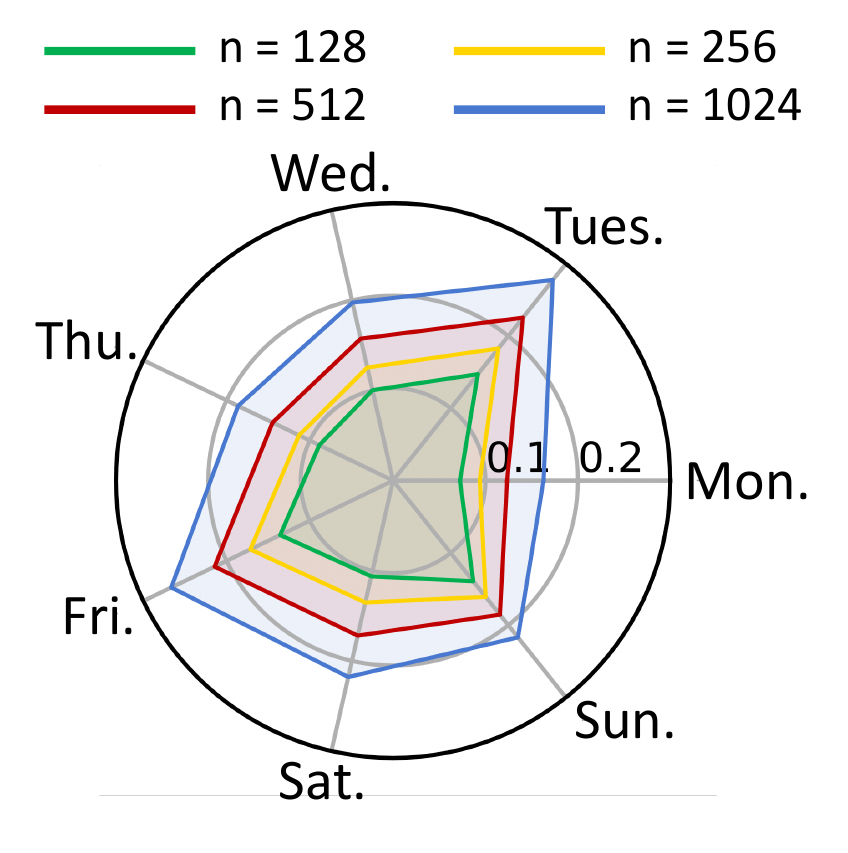}
	}
    \end{center}
    \vspace{-2mm}
    \caption{\textbf{Flow length distribution and entropy ratios.}} 
    \vspace{-4mm}
\label{flow distribution}
\end{figure}

\noindent \textbf{Flow Classification.} \model takes sequences of fixed length as input. Typically, the input length is a power of $2$, and we assume it to be $n$, where $n = 2 ^ k$. Figure~\ref{fig:flow length distribution} shows the probability density function (PDF) and cumulative distribution (CDF) for flow lengths in the MAWI internet traffic dataset (Jun. 2023)~\cite{WIDE2023}. We randomly sample over $1e7$ flows to plot the figure. Clearly, the distribution of flow lengths exhibits a long-tail pattern, with short flows dominating the majority. We obtain the 99th percentile from the cumulative distribution and select the closest $n$ as our hyperparameter for fixed length. Nonetheless, studies of flow distributions~\cite{TOCS03-FlowDistribution} indicate that long flows hold most of the packets. Figure~\ref{fig:entropy distribution} shows the bytes retained for different fixed truncations (i.e., $n$), which can be approximately considered as information entropy, as a proportion of the total bytes of the flow. We randomly sample one week and analyze the complete flow data daily, finding that the information entropy ratios for common values of $n$ do not exceed $0.27$. To address this, we apply two complementary strategies:

\begin{itemize}
    \item \textit{\textbf{Short Flow Padding.}} If $m \leq n$, we append $n - m$ special padding tokens (i.e., $[PAD]$) to the end of its feature sequence.

    \item \textit{\textbf{Long Flow Chunking.}} If $m > n$, we divide its feature sequence into $m - n + 1$ segments, with each segment's index with a range of $[i, i + n)$, $0 \leq i \leq m - n$.
\end{itemize}

\begin{figure}[t]
    \subfigcapskip=-1mm
    \vspace{-2mm}
    \begin{center}
    \subfigure [IPD distribution.]{
        \label{fig:IPD distribution}
		\includegraphics[width=0.22\textwidth]{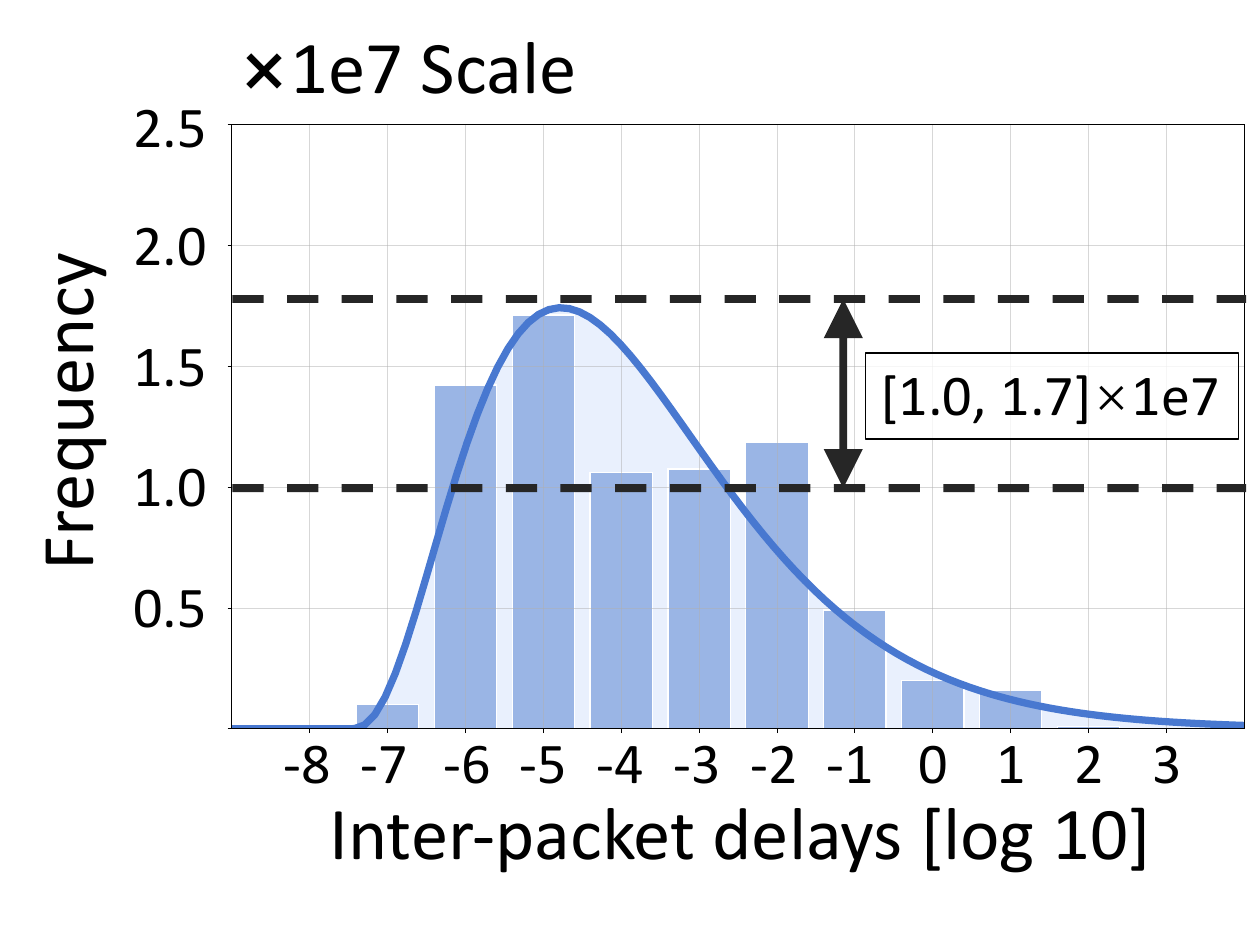}
	}
    \hspace{-3mm}
    \subfigure [Packet size distribution.\textsuperscript{\textbf{1}}]{
        \label{fig:packet size distribution}
		\includegraphics[width=0.22\textwidth]{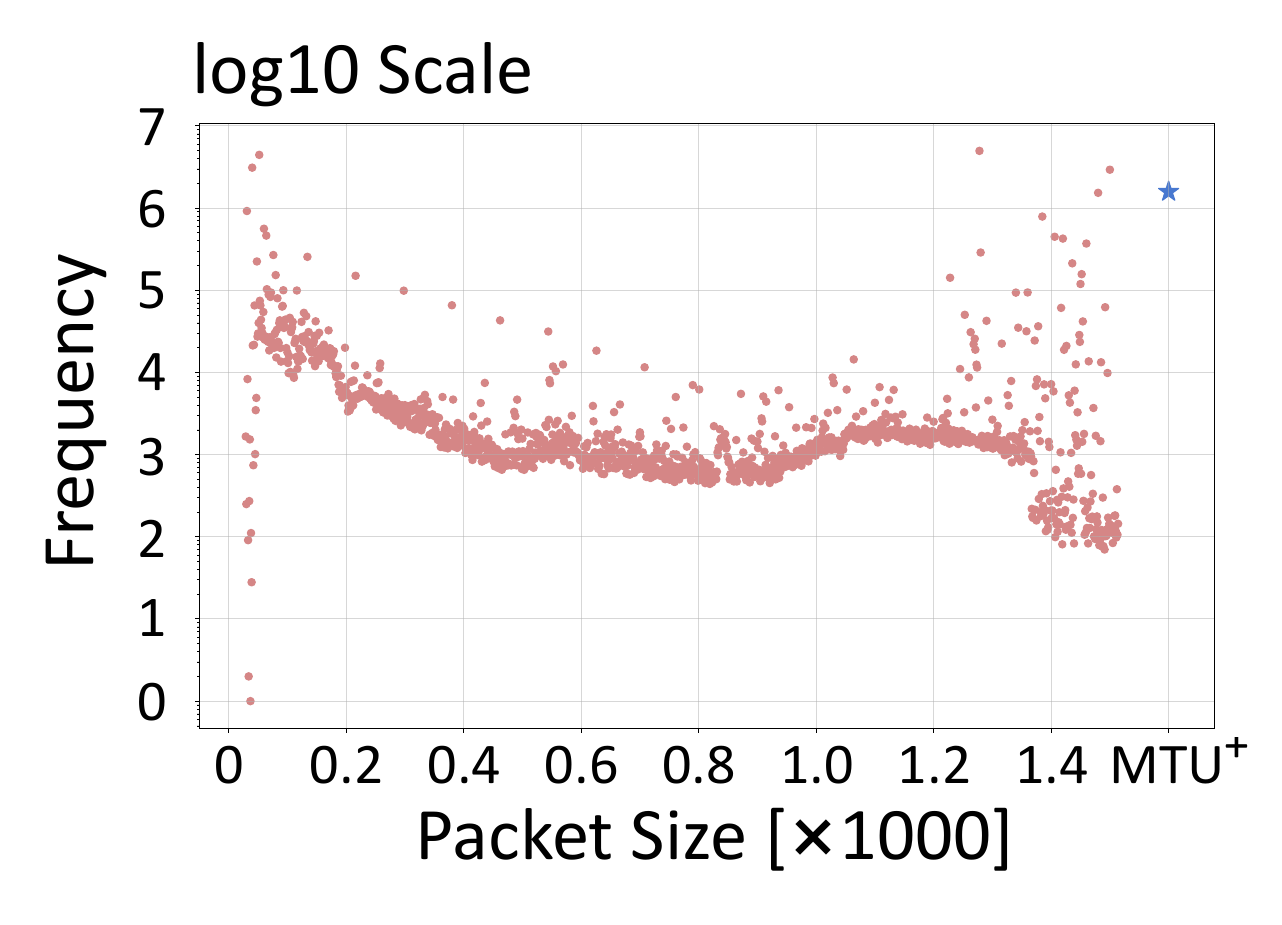}
	}
    \end{center}
    \vspace{-2mm}
    \textsuperscript{1} \small{Size exceeds MTU, due to misconfigurations/specialized protocols.}
    \vspace{-5mm}
    \caption{\textbf{Packet size and IPD distribution.}} 
    \vspace{-6mm}
\label{feature distribution}
\end{figure}

\noindent \textbf{Feature Representation.} Having standardized flow lengths via padding and chunking, we next convert these sequences into discrete tokens that \model can ingest. Specifically, we focus on two per-packet attributes: packet sizes and inter-packet delays (IPDs). To optimize this tokenization process, we study the distribution of packet sizes and IPDs in benign internet traffic. For the IPD feature, we observe that after taking the base-10 logarithm, the data exhibits a more uniform distribution across magnitudes. We randomly sample over $80$ million packets for our analysis and plot these in Figure~\ref{fig:IPD distribution}. The analysis shows that frequencies in the range of $[-6, -2]$ (corresponding to $1e-6$ to $1e-2$ seconds) consistently range between $1.0e7$ and $1.7e7$, while the total for other magnitudes falls below $1e7$. Based on this, we set several logarithmically equal-length intervals and hash the IPDs into the intervals, using the interval indices as the corresponding tokens. We adjust the interval lengths to balance the count of elements within each. The packet sizes exhibit a bimodal distribution: predominantly very short or very long (due to fragmentation), with a more uniform distribution in between, as shown in Figure~\ref{fig:packet size distribution}. Due to its discrete nature, we directly use its value as the token. 
We use a standard Maximum Transmission Unit (MTU) length as the capacity for the packet size vocabulary. This is because we manipulate the traffic on a per-packet basis, making it impossible to generate a single packet that exceeds the MTU. We categorize features significantly exceeding the MTU under a single class and represent them with the $[UNK]$ token. The token vocabularies for packet sizes and IPDs are independent. Moreover, we add special tokens $[PAD]$ and $[MASK]$ to the vocabulary as placeholders and for masking certain tokens, respectively.

We represent each token by two embeddings: token embedding and position embedding. The complete embedding is constructed by summing both of them together.

\begin{itemize}
    \item \textit{\textbf{Token Embedding.}} A high-dimensional vector representing the token, which is randomly initialized and trained jointly with \model.

    \item \textit{\textbf{Position Embedding.}} A vector annotating the token's relative position in the sequence using sinusoidal functions, similar to Transformer~\cite{NIPS17-Transformer}. For chunked features, apart from the first segment, the indices of other segments do not start from $0$. This helps the model learn long flow representations more effectively.
\end{itemize}

\subsection{\model}

Although transformer-based architectures have been applied to network traffic modeling~\cite{WWW22-ET-BERT,SP24-Trafficformer}, most existing BERT-like models are constrained to handling only a single sequence of discrete tokens~\cite{NAACL-HLT18-BERT, Arxiv19-RoBERTa}. In contrast, benign network flows typically involve multiple feature sequences, whose interactions are crucial for capturing real-world traffic patterns. Therefore, the first challenge lies in how to effectively model these interwoven modal features without increasing computational overhead. The second challenge is how to define the \model's training task that enables the model to learn representations of these features directly applicable to adversarial traffic generation, without additional training costs. To address these challenges, we propose a novel bi-cross attention mechanism for efficient fusion of multi-feature data and design a Mask-Fill task to guide \model in acquiring the ability to fill benign traffic packets based on flow context.

\begin{figure}[t]
    \centering
    \includegraphics[width=0.45\textwidth]{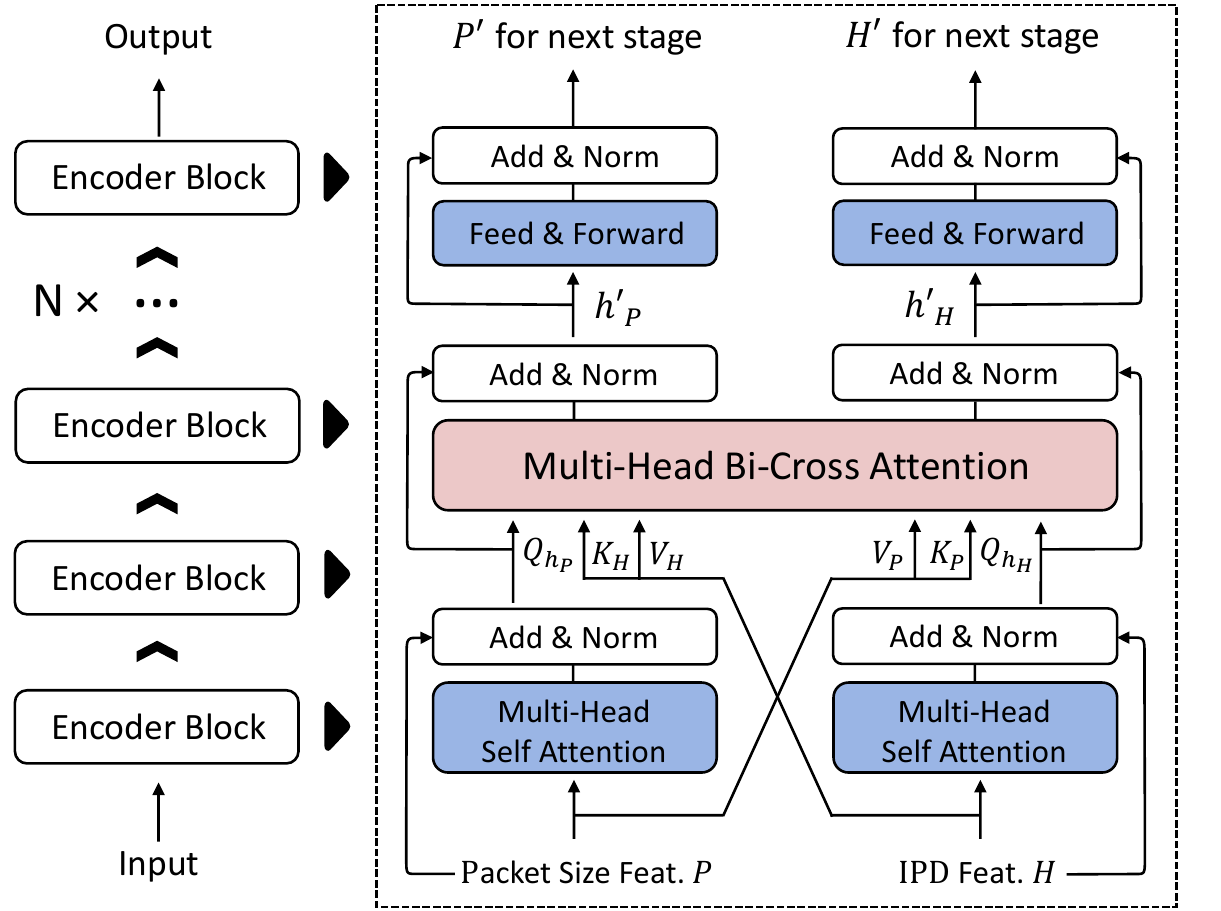}
    \vspace{-2mm}
    \caption{\textbf{Core design of \model.}}
    \label{fig: Core design of Traffic-BERT}
    \vspace{-5mm}
\end{figure}

\noindent \textbf{\model Structure.} The core innovation of \model is the introduction of a bi-cross attention layer within a stack of bidirectional encoder blocks, which is shown in Figure~\ref{fig: Core design of Traffic-BERT}. Each block comprises three principal components: (i) self-attention, (ii) bi-cross attention, and (iii)~a feed-forward network (FFN), with residual connections linking these layers. Generally, the attention mechanism~\cite{NIPS17-Transformer} can be represented as:
\begin{equation}
    \mathcal{A}\textit{ttn.}(Q, K, V) = \text{softmax}\left(\frac{QK^T}{\sqrt{d_k}}\right)V,
\end{equation}
where $Q$, $K$, and $V$ represent the Query, Key, and Value, respectively. They are three sets of linear transformations derived from the packet size and inter-packet delay features. The parameter $\sqrt{d_k}$ represents the dimension of the Key.

Each encoder in \model takes size features $P$ and IPD features $H$ as input, which are first processed by the self-attention layer to yield respective hidden states $h_P$ and $h_H$. Formally:
\begin{align}
    h_P &= P + \mathcal{A}\textit{ttn.}(Q_P, K_P, V_P), \nonumber\\ 
    h_H &= H + \mathcal{A}\textit{ttn.}(Q_H, K_H, V_H).   
\end{align}

Self-Attention is characterized by deriving both the Query and Key from the same sequence, thereby computing the relative importance of each element with respect to all other elements in that sequence. Next, $h_P$ and $h_H$ are fed into the bi-cross attention layer to compute the interrelated attention features, that is, using one feature sequence to query another feature sequence, and vice versa. This can be formulated as:
\begin{align}
h'_P &= h_P + \mathcal{A}\textit{ttn.}(Q_{h_P}, K_H, V_H), \nonumber\\
h'_H &= h_H + \mathcal{A}\textit{ttn.}(Q_{h_H}, K_P, V_P).
\end{align}

Unlike self-attention, bi-cross attention uses the hidden states $h_P$ and $h_H$ as the Query to compute similarity with the other sequence's output from the previous block, and assigns attention weights to the other's Value. The bi-cross attention shares the same complexity $O(n^2 d_k)$ with self-attention, providing an efficient solution for multi-feature sequences from distinct feature spaces. It enables the model to better capture the long-term interactions and dependencies between different feature sequences in network traffic, significantly enhancing the semantic understanding of the benign flow.

The outputs from the bi-cross attention layer are then passed through the feed-forward network layer, serving as the input for the next encoder block. The output from the last encoder block is then passed through a linear layer to obtain the probability distribution of the output tokens.

\label{chapter: post-processing}
\noindent \textbf{\model Optimization.} We train \model with a Mask-Fill task that generalizes RoBERTa's Dynamic Masking~\cite{Arxiv19-RoBERTa} to handle multiple correlated feature sequences. In each training step, we select $15\%$ of token positions for masking. When a position is chosen, tokens in both sequences are masked simultaneously: $80\%$ are replaced with $[MASK]$, $10\%$ with a random token, and $10\%$ remain unchanged. This dual-sequence masking scheme not only compels \model to master deep bidirectional semantics within individual feature sequences, but also reinforces the cross-feature interactions introduced by our bi-cross attention. The trained \model can reconstruct realistic per-packet attributes from partial observations and thus can be directly applied to the second stage in \S~\ref{traffic generation}.

In a few high-speed traffic generation cases, we optionally apply the Constrained Decoding mechanism~\cite{NAACL18-ConstraintDecoding}. This mechanism restricts the model's output to a predefined range, ensuring that only tokens within this range are considered. However, in most cases, such explicit constraints are unnecessary because the Mask-Fill task itself already biases \model toward valid, realistic traffic patterns.

\section{Adversarial Traffic Generation}\label{chapter:stage2}
\label{traffic generation}

In this section, we present the technical details of generating adversarial traffic based on deep RL, focusing on its formulation, optimization, and runtime inference. \this addresses four main challenges:

\begin{enumerate}[leftmargin=*]
    \item \textbf{Efficient Learning from Limited Feedback.} We carefully design the state space (\S~\ref{formulation}), and leverage \model to guide RL (\S~\ref{optimization}), to achieve efficient exploration under black-box settings.

    \item \textbf{Preserving Malicious Functionality.} We adopt a tailored action space and an effectiveness penalty (\S~\ref{formulation}) to maintain malicious behavior within domain constraints.

    \item \textbf{Reducing Training \& Inference Overhead.} We introduce a dissimilarity penalty (\S~\ref{formulation}) and employ a two-stage decoupled training scheme (\S~\ref{optimization}) to minimize overhead while ensuring stability.

    \item \textbf{Enabling Attacks Without Feedback.} We propose an offline reward estimation mechanism (\S~\ref{inference}) that supports real-world adversarial generation when direct feedback is unavailable during inference.
\end{enumerate}

\subsection{MDP Formulation}
\label{formulation}

\this aims to generate effective adversarial malicious traffic that evades black-box detection systems with minimal modifications. To achieve this, we perturb only two domain-agnostic features learned by \model from benign traffic: packet sizes and inter-packet delays. We incorporate specific penalty terms to maintain malicious effectiveness and adhere to domain constraints. This leads to the following definition of our adversarial traffic generation process.

\noindent \textbf{Definition 1 (Adversarial Traffic Generation)}. Given an ML-based malicious traffic detection system $f \in \mathcal{F}$ and an instance $x$ from the malicious traffic space $\mathcal{X}$, the objective of the adversarial traffic generator $G(\cdot, \cdot) : \mathcal{F} \times \mathcal{X} \to \mathcal{X}$ can be defined as:
\hypertarget{def1}{
\begin{align}
\underset{\tilde{x}}{\mathrm{argmax}} \quad & \mathbb{I}(f(\tilde{x}) \neq f(x)) \nonumber\\
\text{s.t.} \quad & \tilde{x} = G(f, x), \nonumber\\
& D(\tilde{x}, x) \leq \tau, \nonumber\\
& M(\tilde{x}, x, \mathcal{X}) = 1,
\end{align}
}
\noindent where $ D(\cdot, \cdot): \mathcal{X} \times \mathcal{X} \to \mathbb{R}^{+} $ is a distance metric that measures the dissimilarity between the adversarial traffic and the original malicious traffic, ensuring that the perturbed instance $\tilde{x}$ is close to the original instance $x$ within the threshold $\tau$, and $M(\cdot, \cdot, \mathcal{X}): \mathcal{X} \times \mathcal{X} \to \{0, 1\}$ is an equivalence indicator that indicates whether the perturbed instance $\tilde{x}$ is equivalent in effectiveness to the original malicious instance.

Note that this optimization problem is difficult to address directly because the objective and the malicious equivalence constraint $M$ are binary and non-differentiable functions, and the distance metric $D$ depends on the way adversarial traffic $\tilde{x}$ is generated. However, we can overcome these challenges by leveraging RL. To do this, we model the attack procedure as a Finite Horizon Markov Decision Process $MDP=(\mathcal{S},\mathcal{A},\mathcal{P},\mathcal{R},\mathcal{T})$. The definition of such an MDP is as follows:

\noindent \textbf{State Space} ($\mathcal{S}$): The state $s_{t} \in \mathcal{S}$ at time $t$ is represented by the tuple $(P_{t}, H_{t})$, where $P_{t} = [p_{0, t}, p_{1, t}, \ldots, p_{n, t}]$ represents the sequence of packet sizes at time $t$ and $H_{t} = [h_{0, t}, h_{1, t}, \ldots, h_{n, t}]$ represents the sequence of inter-packet delays at time $t$. The initial state $s_{0} = (P_{0}, H_{0})$ represents the features extracted from the original malicious traffic.

\noindent \textbf{Action Space} ($\mathcal{A}$): The attacker is allowed to modify the features of a single packet or insert a chaff packet (i.e., an intentionally injected, non‑functional packet) into the flow per step. Therefore, for each feature sequence within the state $s_t$ of length $n$, the size of the action space is $2n + 1$, where each action $a_t \in \mathcal{A}$ represents the index of the modification or insertion. Specifically, when $a_t$ is odd, it indicates the attacker's intention to modify the element at position $\lfloor a_t / 2 \rfloor$ in each sequence; when $a_t$ is even, it indicates the intention to add a new element at position $a_t / 2$ in each sequence. \this does not perturb existing payloads to maintain traffic functionality and avoid introducing detectable artifacts.

\noindent \textbf{Transition Probabilities} ($\mathcal{P}$): Our environment is deterministic. That is, for any given state $s_t$ and action $a_t$, the state transition probability $\mathcal{P}$ can be represented as:
\begin{align}
\mathcal{P}(s_{t+1} | s_t, a_t) = 
\begin{cases} 
1 & \text{if } s_{t+1} = \textit{Trans}(s_t, a_t), \\
0 & \text{otherwise}.
\end{cases}
\end{align}

\noindent \textbf{Reward Function} ($\mathcal{R}$): The reward function is a composite of three distinct components, formalized as
\hypertarget{reward}{
\begin{align}
r(s_t, a_t) = r_E(s_t, a_t) + \beta \cdot r_D(s_t, a_t) + \gamma \cdot r_M(s_t, a_t),    
\end{align}
}
where $\beta$ and $\gamma$ are non-negative hyperparameters. 

The term $r_E$ aligns with the optimization objective in \hyperlink{def1}{(4)}, which can be defined as:
\begin{align}
r_E(s_t, a_t) &= \frac{N_{\text{evade}}(s_{t+1}) - N_{\text{evade}}(s_{t})}{N_{\text{total}}},
\end{align}
where $N_\text{evade}(\cdot): \mathcal{S} \to \mathbb{Z}_{0}^{+}$ represents the number of non-chaff packets that evade the target, and $N_\text{total}=n$ represents the total number of packets in the original malicious traffic.

The term $r_D$ is the dissimilarity penalty which is derived from the distance metric $D(\cdot, \cdot)$. In our scenario, we use the \textit{Edit Distance} between the current state $s_t$ and the previous state $s_{t-1}$. Note that since the attacker modifies or inserts exactly one packet at each step, we have: 
\begin{align}
r_D(s_t, a_t) = -1.
\end{align}

$r_D$ serves to minimize the distance between the adversarial and original malicious traffic, ensuring that \this achieves its adversarial objectives in as few steps as possible, as each additional step incurs a non-positive reward. In general, this design preserves the stealthiness of adversarial traffic and reduces the probability that the  perturbations are detected. On the other hand, \this achieves its goals with fewer modifications, thereby accelerating inference speed.

The term $r_M$ is the effectiveness penalty that depends on the specific intent of the attack. For instance, for DoS traffic, $r_M$ can be defined as the rate of the traffic flow. Conversely, for maliciousness that stems from the payload, such as phishing traffic, $r_M$ can be set to zero, as our adversarial traffic generation process does not impact the payload.

\noindent \textbf{Horizon} ($\mathcal{T}$): The process terminates in either of two situations: first, in the step $t = \tau$, which is consistent with the constraint $D(\tilde{x}, x) \leq \tau$ in \hyperlink{def1}{(1)} as a measure of the maximum permissible distance between the adversarial and original malicious traffic; second, when the reward $r_E(s_t, a_t) > \xi$, indicating a successful evasion ratio greater than the threshold $\xi$. This dual-condition criterion guarantees a bounded process. 

\subsection{Policy Optimization}
\label{optimization}

Algorithm~\ref{algorithm1} shows the process of training \this. Given the MDP setup as defined in \S~\ref{formulation}, a sampled MDP trajectory will be $(s_0, a_0, r_0, s_1, a_1, r_1, \ldots, s_{\tilde{t}}, a_{\tilde{t}}, r_{\tilde{t}})$, where $\tilde{t} \leq \tau$. To handle the problem's large discrete action space, we employ the Soft Actor-Critic (SAC) algorithm for optimization, which is well known for strong exploration capabilities.

The SAC algorithm is an off-policy maximum entropy RL method, aiming to maximize the balance between expected return and entropy, where the entropy signifies the randomness of the policy. Its objective function is given by:
\begin{align}
    \pi^* = \arg\max_{\pi} \mathbb{E}_{\pi} \left[ \sum_{t} \eta^t \left( r(s_t, a_t) + \alpha \mathcal{H}(\pi(\cdot | s_t)) \right) \right],
\end{align}
where $\pi$ is the stochastic policy to be optimized, $\alpha$ is the temperature hyperparameter that controls the trade-off between exploration and exploitation, $\eta$ represents the discount factor, and the entropy of the policy $\mathcal{H}(\pi(\cdot | s_t))$ is defined as the expected value of the negative log-probability of the actions taken according to the policy:
\begin{align}
    \mathcal{H}(\pi(\cdot | s_t)) = \mathbb{E}_{a_t \sim \pi(\cdot | s_t)}[-\log \pi(a_t | s_t)].
\end{align}

Given the optimization objective, we build a policy network to approximate the optimal policy $\pi^*$, for which we employ Gated Recurrent Units (GRUs) as the backbone. We choose GRUs for two reasons: on the one hand, as a classical type of Recurrent Neural Network (RNN), GRUs are capable of understanding the semantics within the traffic feature sequences; on the other hand, compared to the more computationally demanding \model, GRUs offer a balance between complexity and performance, enhancing the training efficiency of the reinforcement learning model.

In each step $t$, the policy network takes as input the concatenated feature sequences of packet sizes and IPDs at state $s_t$, and outputs a distribution over the action space $\mathcal{A}$, as detailed in \S~\ref{formulation}. An action $a_t$ is then sampled from this distribution. Notably, when $a_t$ is odd, the $\lfloor a_t / 2 \rfloor$-th element of the inter-packet delay sequence of state $s_t$ is replaced with a $[MASK]$ token, indicating the attacker's intent to modify the transmission timestamp of the packet at that position. Consequently, the state $s_t$ is transformed into 
\begin{align}
    &s'_{t} \triangleq (P'_{t}, H'_{t}) = \left([p_{0, t}, p_{1, t}, \ldots, p_{n, t}], \right. \nonumber\\
    &\left. [h_{0, t}, h_{1, t}, \ldots, h_{\lfloor a_t / 2 \rfloor-1, t}, [MASK], h_{\lfloor a_t / 2 \rfloor + 1, t}, \ldots, h_{n, t}]\right).
\end{align}

In this context, the packet size sequence remains unaltered as changes to packet sizes might violate the domain constraints of the packet. Correspondingly, when $a_t$ is even, a $[MASK]$ token is inserted at the $a_t / 2$ position of both feature sequences, indicating the attacker's intention to insert a chaff packet at that position. In this case, the state $s_t$ is transformed into
\begin{align}
s'_{t} &\triangleq (P'_{t}, H'_{t}) \nonumber\\
&= \left([p_{0, t}, p_{1, t}, \ldots, p_{a_t / 2 - 1, t}, [MASK], p_{ a_t / 2 , t}, \ldots, p_{n, t}], \right. \nonumber\\
&\left. [h_{0, t}, h_{1, t}, \ldots, h_{ a_t / 2 - 1, t}, [MASK], h_{a_t / 2, t}, \ldots, h_{n, t}]\right).
\end{align}

\begin{algorithm}[t]
\caption{\this Training Process}
\label{algorithm1}
\begin{algorithmic}[1]

\State Initialize policy network $\pi_\phi(a|s)$, Q-networks $Q_{\omega_1}(s,a)$, $Q_{\omega_2}(s,a)$, and experience replay buffer $\mathcal{B}$

\For{each iteration}
    \State Sample a malicious flow and get initial state $s_0$
    \For{each environment step $t$}
        \State Observe state $s_t$ and select action $a_t \sim \pi_\phi(\cdot|s_t)$ based on the current policy
        \State Modify $s_t$ by inserting $[MASK]$ or replacing features with $[MASK]$ to produce $s'_t$
        \State Use \model for Mask-Fill task to fill $[MASK]$, obtaining $s_{t+1}$
        \State Restore $s_{t+1}$ to adversarial malicious traffic
        \State Send the adversarial malicious traffic and compute reward $r_t = r_E + \beta \cdot r_D + \gamma \cdot r_M$
        \State Store transition tuple $(s_t, a_t, r_t, s_{t+1})$ in $\mathcal{B}$
        \If{${|\mathcal{B}|}$ exceeds minimum replay buffer size}
            \State Sample mini-batch $\{s_{\bar{t}}, a_{\bar{t}}, r_{\bar{t}}, s_{{\bar{t}}+1}\}$ from $\mathcal{B}$
            \State Compute target value for each Q-network: 
            \Statex $y_{\bar{t}} = r_{\bar{t}} + \eta\min_{i=1,2}Q_{\bar{\omega}_i}(s_{{\bar{t}}+1}, \pi(\cdot|s_{{\bar{t}}+1})) - \alpha \log \pi(a_{\bar{t}}|s_{\bar{t}})$
            \State Update Q-networks:
            \Statex $\omega_i \leftarrow \omega_i - \lambda_{Q}\nabla_{\omega_i}\sum(Q_{\omega_i}(s_{\bar{t}}, a_{\bar{t}}) - y_{\bar{t}})^2$
            \State Update policy network:
            \Statex $\phi \leftarrow \phi - \lambda_{\pi}\nabla_\phi \sum \left(\alpha \log(\pi_\phi(a_{\bar{t}}|s_{\bar{t}})) - Q_{\omega_i}(s_{\bar{t}}, a_{\bar{t}})\right)$
            \State Update target Q-networks:
            \Statex $Q_{\bar{\omega}_i} \leftarrow \lambda Q_{\omega_i} + (1 - \lambda) Q_{\bar{\omega}_i}$
            \State Update entropy temperature $\alpha$:
            \Statex $\alpha \leftarrow \alpha - \lambda_\alpha \nabla_\alpha \sum \left( - \alpha \log(\pi(a_{\bar{t}}|s_{\bar{t}})) - \alpha \mathcal{H}_0 \right)$
        \EndIf
    \EndFor
\EndFor
\end{algorithmic}
\end{algorithm}

Considering that the fixed length $n$ may exceed the actual length of a flow, not all actions are feasible. To address this issue, we employ an Invalid Action Masking mechanism~\cite{FLAIRS22-Huang}, adjusting the probabilities of infeasible actions to a large negative value, and then re-normalizing the probability distribution to ensure the effectiveness of the chosen actions.

Once $s'_t$ is obtained, the attacker leverages \model in the Mask-Fill task to embed benign traffic patterns, thereby deriving the next state $s_{t+1}$. During this process, \model's parameters are fixed and considered a part of the environment. This decoupling of training significantly improves training efficiency.

According to \S~\ref{chapter:threat}, the attacker can conduct a reconnaissance phase to gather pass/fail feedback, i.e., by observing whether there is any response to the attack traffic. Based on this, the attacker restores the adversarial flow from $s_{t+1}$ and sends it to the target. The process is straightforward as \this only modifies timestamps or inserts new chaff packets. These chaff packets share the same source/destination IPs and ports as the malicious flow. Their payloads are randomly populated to match the packet size features. Following prior work~\cite{CoNEXTWorkshop19-AdvNIDS, JSAC21-TM}, for example, we use incorrect sequence numbers for TCP, set a short TTL for UDP packets, or send orphan IP fragments for other protocols which are discarded after a reassembly timeout~\cite{RFC791}. After the traffic is sent, the attacker calculates the reward $r = r_E + \beta \cdot r_D + \gamma \cdot r_M$. Finally, the resulting transition $(s_t, a_t, r_t, s_{t+1})$ is stored in the experience replay buffer $\mathcal{B}$ for further policy optimization.

Given the optimization objective and the transitions, we model two state-action value functions (a.k.a., Q-networks) $Q_{\omega_1}$ and $Q_{\omega_2}$. The utilization of double Q-networks helps to mitigate the overestimation of action values. Following the Soft Bellman Equation~\cite{ICML17-SoftQLearning}, each Q-network can be updated by minimizing the following loss function:
\begin{align}
    \mathcal{L}_Q(\omega) &= \mathbb{E}_{(s_t, a_t, r_t, s_{t+1}) \sim \mathcal{B}}\bigg[\frac{1}{2} \bigg(Q_{\omega}(s_t, a_t) - y_t\bigg)^2\bigg], \nonumber\\
    y_t &= r_t + \eta(\min_{j=1,2} Q_{\bar{\omega}_j}(s_{t+1}, \pi(\cdot|s_{t+1})) \nonumber\\
    &\qquad\qquad\qquad\qquad\quad\ - \alpha \log \pi(a_t|s_t)),
\end{align}
where $Q_{\bar{\omega}}$ represents the target Q-network~\cite{Nature15-DQN}, which helps to smooth the learning updates. The target Q-networks are updated using Q-networks:
\begin{align}
    Q_{\bar{\omega}_i} \leftarrow \lambda Q_{\omega_i} + (1 - \lambda) Q_{\bar{\omega}_i}.
\end{align}

The policy network optimization is achieved by minimizing the Kullback-Leibler Divergence from the exponential of the Q-network. This results in the following loss function:
\begin{align}
    \mathcal{L}_\pi(\theta) = \mathbb{E}_{s \sim \mathcal{B}, a \sim \pi}\left[\alpha \log(\pi(a|s)) - \min_{j=1,2} Q_{\bar{\omega}_j}(s, a)\right].
\end{align}

Also, following~\cite{Arxiv18-SAC}, we employ an automated entropy adjustment mechanism for the temperature parameter $\alpha$:
\begin{align}
    \mathcal{L}(\alpha) = \mathbb{E}_{s_t \sim \mathcal{B}, a_t \sim \pi(\cdot|s_t)}[-\alpha \log \pi(a_t|s_t) - \alpha \mathcal{H}_0].
\end{align}

\subsection{Runtime Inference}
\label{inference}

\begin{algorithm}[t]
\caption{\this Inference Process}
\label{algorithm2}
\begin{algorithmic}[1]

\State Initialize policy network $\pi_\phi(a|s)$ with trained parameters
\State Initialize Q-networks $Q_{\omega_1}(s,a)$ and $Q_{\omega_2}(s,a)$ with trained parameters
\State Set step $t = 0$
\State Transform the malicious flow into initial state $s_0$

\While{$t < \tau$}
    \State Observe state $s_t$ and select action $a_t \sim \pi_\phi(\cdot|s_t)$ based on the policy network
    \State Modify $s_t$ by inserting $[MASK]$ or replacing features with $[MASK]$ to produce $s'_t$
    \State Use \model for Mask-Fill task to fill $[MASK]$, obtaining $s_{t+1}$
    \State Calculate Q-Values 
    \Statex $q_1 \gets Q_{\omega_1}(s_t, a_t)$, $q_2 \gets Q_{\omega_2}(s_t, a_t)$
    \If{$\max_{i=1,2} q_i \geq \xi'$}
        \State \textbf{break} \Comment{Termination condition is met}
    \EndIf
    \State $t \leftarrow t + 1$
\EndWhile
\State Restore $s_{t}$ to the final adversarial malicious traffic

\end{algorithmic}
\end{algorithm}

Algorithm~\ref{algorithm2} shows the inference process of \this. Unlike the training phase, the attacker might not be able to receive reward feedback $r_E$ from the target detection system during the inference phase, which prevents direct evaluation of the termination time for adversarial traffic generation. To address this, we approximate the expected total reward for each action using the maximum value from the two Q-networks. The termination condition for the inference phase of \this is as follows:
\begin{align}
    (t \geq \tau) \lor \left(\max_{i=1,2} Q_{\omega_i}(s_t, a_t) \geq \xi - \beta \cdot r_{D} - \gamma \cdot r_{M}\right),
\end{align}
where the threshold $\xi$ is determined empirically from the training phase. Specifically, we monitor the attack success rate (ASR) during training. Once the ASR stabilizes at a high level, indicating a successfully trained agent, the corresponding Q-value is recorded to serve as the threshold. In cases where the attacker cannot compute $r_M$, the termination condition can be transformed into:
\begin{align}
    (t \geq \tau) \lor \left(\max_{i=1,2} Q_{\omega_i}(s_t, a_t) \geq \xi' \right).
\end{align}

\section{Evaluation}
\label{chapter:evaluation}

\subsection{Experiment Setup}
\label{setup}
\noindent \textbf{Implementation.} Traffic-BERT and the RL pipeline are written in Python v3.8.15 using PyTorch~\cite{Pytorch} v1.13.1. Each adversarial flow produced by \this is delivered over a socket to an Intel DPDK~\cite{DPDK} v24.11.1 worker that emits the actual packets. The DPDK process, written in C (compiled with GCC v9.4.0 -O3 via Meson v0.61.5 and Ninja v1.8.2), pre-allocates NUMA-aware mbuf pools, configures a single 1024-descriptor TX queue, and relies on TSC-based busy-wait pacing to preserve $\mu$s-level inter-packet spacing, thereby avoiding the NIC's internal burst-coalescing that would otherwise distort the on-wire delay. \this runs on a Dell server equipped with two Intel Xeon Gold 6348 CPUs ($2 \times 28$ cores, $112$ threads) and a single NVIDIA Tesla A100 GPU (driver v530.30.02, CUDA~\cite{CUDA} v12.1) under Ubuntu v18.04.6 (Linux 5.4.0-150-generic). The DPDK worker interfaces with an Intel 82599SE NIC  ($2 \times 10 $ Gb/s SFP+ ports). All hyper-parameters are listed in Table~\ref{tab: parameters} in the Appendix.

\noindent \textbf{Datasets.} 
We use real-world backbone network traffic traces from Samplepoint-F of the WIDE MAWI project~\cite{WIDE2023}, collected in June and August 2023, as background traffic. Following established practices~\cite{IMC19-kopp,IMC19-Richter}, we remove scanning traffic that attempts to connect to more than 10\% of IP addresses and apply additional filtering rules~\cite{IMC19-kopp} to eliminate flooding traffic. We then employ the resulting background traffic in two ways: (i) to train \model, using more than 1 million flows collected in June 2023, and (ii) to supplement the target system's training data with flows from August 2023 when the proportion of benign traffic in the malicious dataset is insufficient (Botnet Attacks, see Table~\ref{tab: dataset details}). Notably, this choice does not compromise the black-box setting, as there is no correlation between the distributions of the datasets.

To closely mirror real-world scenarios and highlight NetMasquerade's task-agnostic capabilities, we replay $4$ groups of attacks from multiple datasets, totaling $12$ attacks: (i) Reconnaissance and Scanning Attacks, including host-scanning and fuzz-scanning traffic; (ii) Denial of Service Attacks, covering SSDP and TCP SYN flood traffic; (iii) Botnet Malwares, featuring $4$ common botnet strains—Mirai, Zeus, Storm, and Waledac; and (iv) Encrypted Web Attacks, encompassing webshell, XSS, CSRF, and encrypted spam traffic. The details of the datasets can be found in the Appendix~\ref{subsection: details of datasets}.

\noindent \textbf{Target Systems.} We deliberately select as attack target $6$ existing malicious traffic detection systems that reflect diverse designs. We use $3$ advanced traditional ML-based detection systems: Whisper~\cite{CCS21-Whisper}, FlowLens~\cite{NDSS21-FlowLens}, NetBeacon~\cite{Security23-NetBeacon} and $3$ top-performing DL-based systems: Vanilla feature + RNNs, CICFlowMeter~\cite{ICISSP16-Gil,ICISSP17-Lashkari} + MLP, Kitsune~\cite{NDSS18-Kitsune}. They operate using different learning approaches like supervised classification~\cite{Security23-NetBeacon} and unsupervised anomaly detection~\cite{NDSS18-Kitsune, CCS21-Whisper}, at both the flow-level~\cite{NDSS21-FlowLens} and packet-level~\cite{NDSS18-Kitsune}, and their implementations span both software~\cite{NDSS18-Kitsune, CCS21-Whisper} and programmable switches~\cite{Security23-NetBeacon, NDSS21-FlowLens}. More information about malicious detection systems can be found in Appendix~\ref{subsection: details of target systems}.

\newcommand{\bs}[1]{\color[rgb]{0.117, 0.447, 0.999}#1}
\newcommand{\ws}[1]{\color[rgb]{0.753,0,0}#1}

\renewcommand{\arraystretch}{1.15}
\begin{table*}[!t]
    \footnotesize
    \centering
    \setlength\tabcolsep{1.9pt}
    \caption{Attack Success Rate (ASR) of NetMasquerade and Baselines on Detection Systems}
    \vspace{0mm}
    \resizebox{\textwidth}{!}{
    \begin{threeparttable}
        \begin{tabular}{@{}c|c|c|cc|cc|cccc|cccc|c@{}}
             \toprule
             \multicolumn{2}{c|}{\multirow{2}{*}{Target System}}
             & \multirow{2}{*}{Methods}
             & \multicolumn{2}{c|}{Recon.\&Scan.}
             & \multicolumn{2}{c|}{DoS}
             & \multicolumn{4}{c|}{Botnet}
             & \multicolumn{4}{c|}{Encrypted Web Attacks}
             & \multirow{2}{*}{Overall}
            \\
            \cline{4-5}\cline{6-7}\cline{8-11}\cline{12-15}
             \multicolumn{2}{c|}{}
             &
             & Scan
             & Fuzz.
             & SSDP
             & SYN
             & Mirai 
             & Zeus 
             & Storm 
             & Waledac
             & Webshell
             & XSS
             & CSRF
             & Spam
             & 
            \\
            \midrule
            \multirow{15}{*}{\tabincell{c}{Traditional\\ML-based\\ systems}}                                                                                   
                                          & \multirow{5}{*}{Whisper} & R.M.   & \ws {-}~\tnote{1} & \ws 0.0100  & \ws {-}    & \ws 0.2552  & \ws 0.2324 & \ws 0.1011 & 0.2289     & \ws 0.0585  & \ws 0.0812 & \ws 0.0721 & \ws 0.1717 & \ws 0.0927  & \ws 0.1087 \\
                                          &                          & M.I.   & 0.8907        & 0.0756          & 0.1132     & 0.3346      & 0.5521     & 0.5719     & 0.4590     & 0.4251      & 0.6802     & 0.7010     & 0.7259     & 0.7319  & 0.5218 \\
                                          &                          & T.M.   & 0.9344        & 0.9270          & 0.7712     & 0.2790      & 0.6355     & 0.2551     & \ws 0.1820 & 0.3664      & 0.5839     & 0.5527     & 0.6055     & 0.9072  & 0.5833 \\
                                          &                          & Amoeba & \bs 0.9999    & 0.9934          & \bs 0.9999 & \bs 0.9998  & 0.9167     & 0.9254     & 0.9844     & 0.8970      & \bs 0.9999 & \bs 0.9999 & 0.9966     & 0.8381  & 0.9626\\
                                          &                          & \textbf{NetM.}  & \bs 0.9999    & \bs 0.9965      & \bs 0.9999 & 0.9467      & \bs 0.9988 & \bs 0.9972 & \bs 0.9999 & \bs 0.9355  & \bs 0.9999 & \bs 0.9999 & \bs 0.9999 & \bs 0.9795  & \bs 0.9878\\
            \cline{2-16}
                                          & \multirow{5}{*}{FlowLens}& R.M.   & \ws {-}       & \ws {-}    & \ws 0.1782  & 0.7660      & \ws 0.6893 & \ws 0.0760  & \ws 0.3846  & \ws 0.0434   & \ws 0.0100 & \ws {-}     & 0.0150     & \ws {-} & \ws 0.1802 \\
                                          &                          & M.I.   & 0.9800        & 0.1158     & 0.2375      & \ws 0.5950   & 0.9370    & 0.4941      & 0.6510      & 0.3114       & 0.6391     & 0.5959      & 0.6633     & 0.1313 & 0.5293 \\
                                          &                          & T.M.   & 0.0222        & 0.1525     & 0.9344      & 0.9125      & 0.8591    & 0.2670      & 0.8374      & 0.2899       & 0.0760     & 0.0736      & \ws 0.0036 & 0.3913 & 0.4016 \\
                                          &                          & Amoeba & 0.9976        & \bs 0.9442 & \bs 0.9999  & 0.9990      & 0.8776    & 0.8665     & 0.9252     & 0.8000      & \bs 0.9990 & \bs 0.9999  & 0.9295       & \bs 0.9700 & 0.9424 \\
                                          &                          & \textbf{NetM.}  & \bs 0.9999    & 0.9335     & \bs 0.9999  & \bs 0.9995  & \bs 0.9537 & \bs 0.9102 & \bs 0.9990 & \bs 0.9955  & 0.9795     & \bs 0.9999  & \bs 0.9428   & 0.9475 & \bs 0.9717\\
            \cline{2-16}
                                          & \multirow{5}{*}{NetBeacon}& R.M.  & \ws {-}       & \ws {-}    & 0.5291       & \ws 0.1823  & \ws 0.2864 & \ws 0.0230 & \ws {-}  & 0.0790  & 0.6294 & \ws 0.3916 & \ws 0.1066     & \ws 0.1030              & \ws 0.1942\\
                                          &                          & M.I.   & 0.6511        & \ws {-}    & \ws 0.2285   & 0.2841      & 0.5544     & 0.3455     & 0.3032    & \ws {-}     & 0.8781  & 0.7010     & 0.6446     & 0.1134                 & 0.3920     \\
                                          &                          & T.M.   & 0.6494        & 0.2435     & 0.8577       & 0.4393      & 0.3047     & 0.1992     & 0.4415   & 0.2180      & \ws 0.4585      & 0.5645     & 0.5294     & 0.9091          & 0.4846     \\
                                          &                          & Amoeba & 0.9900       & \bs 0.9999     & 0.9987       & \bs 0.9999  & \bs 0.9999 & 0.5905     & 0.6916   & 0.9727   & 0.9550  & \bs 0.9999     & 0.9894     &  \ws N/A~\tnote{2}  & 0.8490          \\
                                          &                          & \textbf{NetM.}  & \bs 0.9999        & \bs 0.9999     & \bs 0.9999   & \bs 0.9999  & 0.9899     & \bs 0.9449  & \bs 0.9965 & \bs 0.9999  & \bs 0.9999  & \bs 0.9955 & \bs 0.9999 & \bs 0.8448  &  \bs 0.9809\\
            \midrule
            \multirow{15}{*}{\tabincell{c}{DL-based \\ systems}} 
                                          & \multirow{5}{*}{Vanilla} & R.M.   & \ws {-}       & 0.3660      & 0.0455        & 0.5815      & 0.1163       & \ws {-}    & \ws {-}    & 0.3299     & \ws 0.0118 & \ws {-}    & \ws 0.0050 & \ws 0.0515 & \ws 0.1256\\
                                          &                          & M.I.   & 0.9510         & \ws {-}    & \ws {-}       & \ws 0.3355  & 0.8769       & \ws {-}    & 0.5415      & 0.6711     & 0.6085     & 0.5353     & 0.6751     & 0.1958     & 0.4492 \\
                                          &                          & T.M.   & \ws {-}       & 0.0375      & 0.8600        & 0.6550      & \ws 0.0790   & 0.2232     & 0.2595      & \ws 0.1617 & 0.0492     & 0.0278     & 0.0264     & \bs 0.8636 & 0.2702 \\
                                          &                          & Amoeba & \bs 0.9999    & \bs 0.9999 & \bs 0.9999    & \bs 0.9999  & 0.8038       & 0.7156     & 0.6540     & 0.2682    & 0.9975     & \bs 0.9999 & 0.9455     & 0.2538     & 0.8032 \\
                                          &                          & \textbf{NetM.}  & \bs 0.9999    & 0.9985     & 0.9825        & 0.9890      & \bs 0.9817   & \bs 0.9894 & \bs 0.9805  & \bs 0.9687 & \bs 0.9999 & \bs 0.9999 & \bs 0.9999  & 0.8485    & \bs 0.9782\\
            \cline{2-16}
                                          & \multirow{5}{*}{CIC.}    & R.M.   & \ws {-}       & 0.0422     & \ws 0.1100   & 0.6398       & \ws 0.5578  & 0.2467     & \ws 0.2922 & \ws 0.0301 & \ws 0.0151 & \ws 0.1855 & 0.4467     & 0.1031  & \ws 0.2224 \\
                                          &                          & M.I.   & 0.2300        & 0.1367     & 0.9711       & \ws 0.5735   & 0.7111      & 0.3956     & 0.5396     & 0.2122     & 0.7011     & 0.6185     & 0.6598     & 0.2886  & 0.5032 \\
                                          &                          & T.M.   & 0.1444        & \ws {-}    & 0.9822       & 0.6520       & 0.6656      & \ws 0.1433 & 0.3026     & 0.1021     & 0.0311     & 0.0445     & \ws 0.3381 & 0.6391  & 0.3371 \\
                                          &                          & Amoeba & \bs 0.9999    & \ws N/A    & \bs 0.9999   & 0.9112       & 0.9980      & \bs 0.9999 & 0.8704     & 0.8182     & \bs 0.9800 & 0.9865     & \bs 0.9999 & \ws N/A & 0.7970\\
                                          &                          & \textbf{NetM.}  & \bs 0.9999    & \bs 0.9744 & \bs 0.9999  & \bs 0.9959   & \bs 0.9999 & 0.9867     & \bs 0.8898 & \bs 0.9810 & 0.9767     & \bs 0.9999 & \bs 0.9999 & \bs 0.7475  & \bs 0.9626\\
            \cline{2-16}
                                          & \multirow{5}{*}{Kitsune} & R.M.   & \ws {-}       & \ws {-}    & 0.2379        & 0.3744   & \ws 0.2949     & \ws 0.0360      & 0.0990      & \ws 0.2901  & \ws {-} & \ws0.0277   & \ws 0.0374         & \ws {-} & \ws 0.1165 \\
                                          &                          & M.I.   & 0.3514         & 0.4484      & \ws 0.0913  & \ws 0.1815 & 0.8109     & 0.0801       & \ws 0.4424  & 0.6334   & 0.6159  & 0.4498      & 0.3493              & 0.5359         &  0.4159 \\
                                          &                          & T.M.   & \bs 0.9760     & \bs 0.9860  & 0.7848      & 0.5590   & 0.9049         & 0.4735          & 0.8318     & 0.7878      & 0.8884  & 0.8965    & 0.8406          & 0.6949        & 0.8020 \\
                                          &                          & Amoeba & 0.9339         & \ws N/A    & \bs 0.8949   & 0.9292   & 0.9915         & \bs 0.9449      & 0.7256      & 0.4595      & 0.4355  & 0.7814       &  0.7017              & \ws N/A & 0.6498 \\
                                          &                          & \textbf{NetM.}  & 0.9049         & 0.9850      & 0.8218      & \bs 0.9333   & \bs 0.9968 & 0.9359      & \bs 0.9911  & \bs 0.9291  & \bs 0.9219  & 0.9231       & \bs 0.9177           & \bs 0.7522 & \bs 0.9177 \\
            \bottomrule
        \end{tabular}
    \begin{tablenotes}
        \scriptsize
        \item[1] These methods cannot successfully attack the target (ASR $<0.01$).
        \item[2] These methods cannot generate legitimate traffic (with illegal packet sizes).
    \end{tablenotes}
    \end{threeparttable}
    \label{tab: asr comparison}
    \vspace{0mm}
    }
\end{table*}

\noindent \textbf{Baselines.} To validate \this, we select $2$ classic and $2$ state-of-the-art attack methods as baselines:
\begin{itemize}[leftmargin=*]
    \item Random Mutation. Random Mutation refers to the technique of obscuring traffic by randomly adjusting IPDs. This traditional method has been demonstrated to be powerful in several works~\cite{EAI-ETSS18-Homoliak, WOOT08-Stinson}, and existing attack tools also employ this method~\cite{cobaltstrike, GodzillaGithub}. In our experiments, the randomization of time intervals follows a Gaussian distribution based on the malicious flow's mean and variance. The number of mutated packets matches \this's modification steps. 
    \item Mutate-and-Inject. We combine Random Mutation and Packet Injection to create a comprehensive attack strategy, which has been used as a standard for evaluating the robustness of advanced detection systems~\cite{CCS21-Whisper}. For Random Mutation, we follow the same rules described above. For Packet Injection, we either inject chaff packets with random sizes and intervals into the malicious flow or duplicate segments of the malicious packets. The modification steps match those in \this.
    \item Traffic Manipulator~\cite{JSAC21-TM}. Traffic Manipulator is the SOTA attack algorithm capable of generating practical adversarial traffic against malicious traffic detection systems in a gray-box scenario. Traffic Manipulator learns adversarial traffic patterns with GANs and adjusts the patterns of malicious traffic with the particle swarm optimization algorithm. We use the open-source implementation of Traffic Manipulator~\cite{TrafficManipulatorGithub} and retrain the model. We use the Kitsune Feature Extractor as the default for Traffic Manipulator, following the paper. This means that for Kitsune, it is a gray-box attack, whereas for other systems, it is a black-box attack.
    \item Amoeba~\cite{CoNEXT23-Amoeba}. Amoeba is designed with a per-packet adjustment technique to circumvent censorship in a black-box scenario. Specifically, Amoeba leverages RL to truncate and pad packets, which are then reassembled by the remote end after passing through the censorship system. We disregard the practical applicability of Amoeba's splitting strategy under different traffic conditions and only evaluate whether it can bypass different types of detection systems. We adopt its open-source implementation and retrain the model.
    
\end{itemize}

\noindent \textbf{Metrics.} We use AUC (Area Under the Curve) and F1 score to assess the effectiveness of the malicious traffic detection systems, and attack success rate (ASR) to measure the performance of attack methods. Specifically, ASR measures the fraction of malicious flows that are not detected when the IDS uses the threshold rendering the highest F1 score on the validation data. We also measure the bandwidth (megabits per second, Mbps) of both malicious and adversarial traffic to illustrate the impact of perturbations. Moreover, we measure throughput (packets per second, PPS) to show the efficiency.

\subsection{Attack Performance}

\begin{figure}[t]
    \subfigcapskip=-1mm
    \begin{center}
	\subfigure[NetBeacon.]{
        \hspace{-2mm}
        \label{fig: qvalue advanced}
		\includegraphics[width=0.225\textwidth]{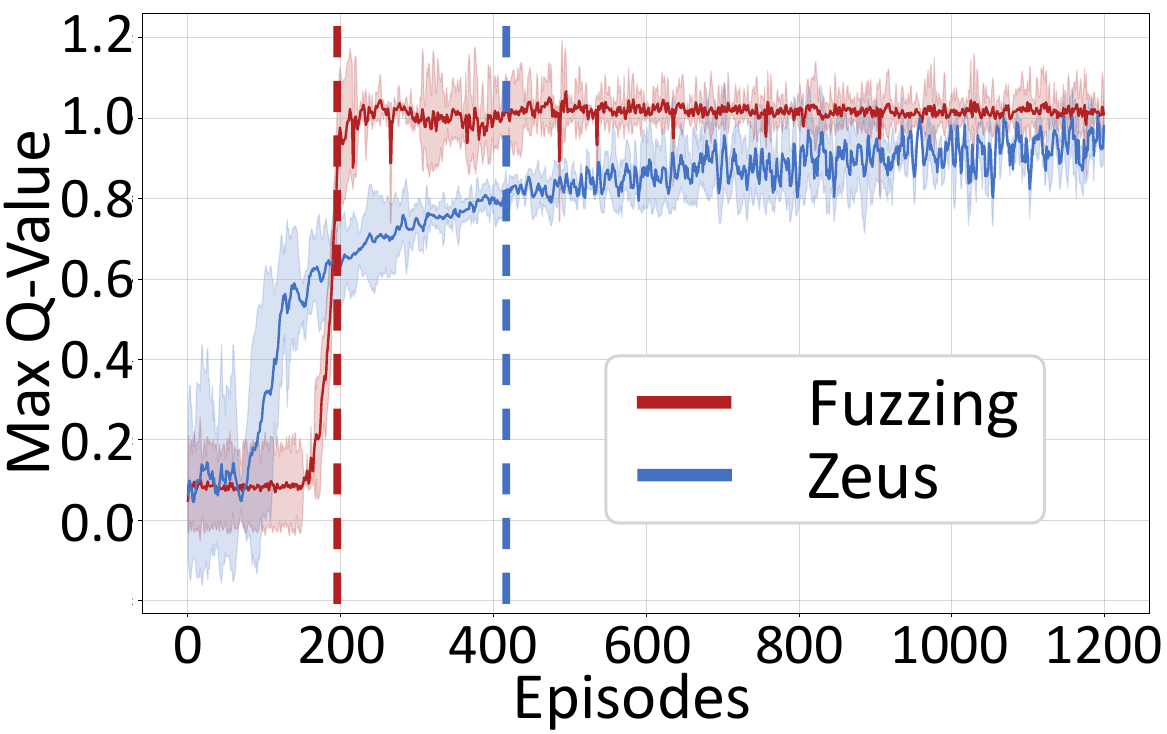}
	}
	\hspace{-2mm}
	\subfigure [CICFlowMeter + MLP.]{
        \hspace{-2mm}
        \label{fig: qvalue classic}
		\includegraphics[width=0.225\textwidth]{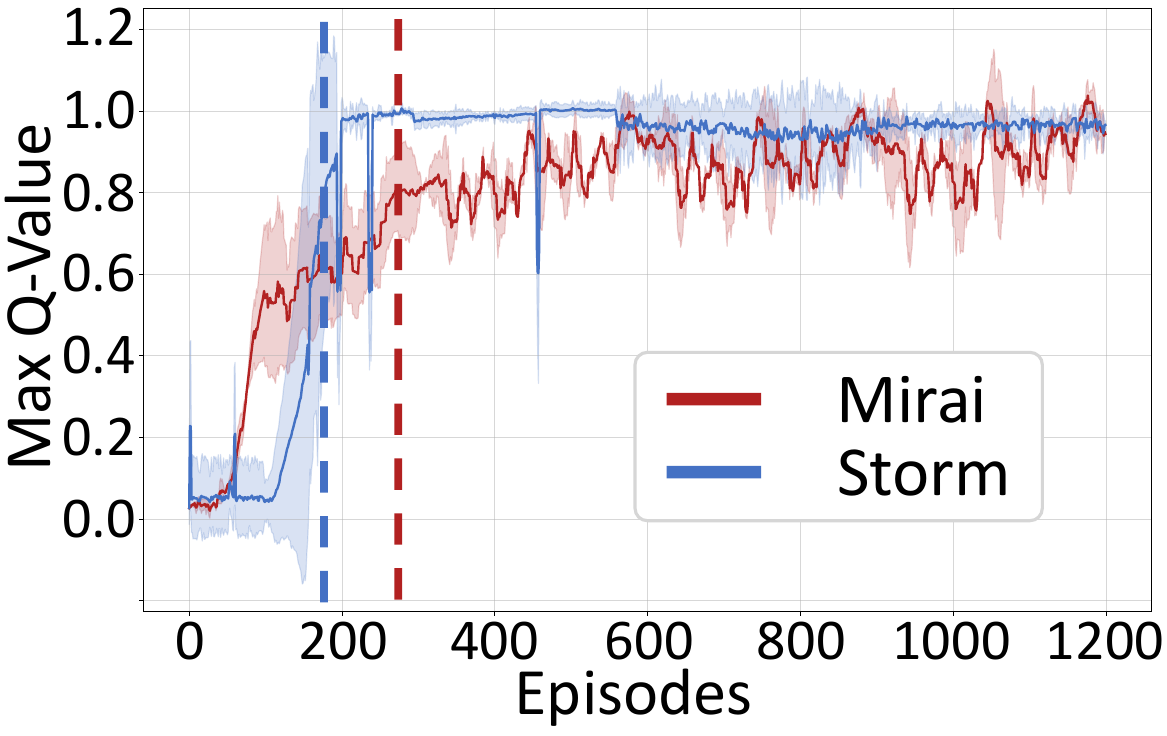}
	}
    \vspace{-2mm}
    \caption{\textbf{Max Q-Value during the training phase.} The shaded region represents a standard deviation of the average evaluation over $5$ trials. Curves are smoothed uniformly for visual clarity.}
    \label{fig: qvalue}
    \end{center}
    \vspace{-8mm}
\end{figure}

Table~\ref{tab: asr comparison} presents the attack success rate against advanced detection systems. \this achieves $0.7475 \sim 0.999$ ASR, with an average of $0.9878$, $0.9717$, $0.9809$, $0.9782$, $0.9626$ and $0.991$ against Whisper, FlowLens, NetBeacon, Vanilla, CICFlowMeter and Kitsune detection systems, showing an improvement of $2.61\%$, $3.11\%$, $15.53\%$, $21.88\%$, $20.78\%$, and $14.42\%$ over the best performance of the baselines. In 56 out of the 72 evaluated scenarios, \this achieves the highest attack success rate. By contrast, Amoeba matches or exceeds \this's performance in 26 scenarios, yet in certain cases its success rate plummets below 30\% and even produces flows with illegal packet sizes. Moreover, Amoeba relies on truncating and padding every single packet to craft adversarial traffic, requiring cooperation from the receiving endpoint to reassemble these packets. As a result, this packet-level manipulation can fail in practical attack scenarios (e.g., spam), where such coordination is typically unavailable. Meanwhile, we observe that Traffic Manipulator's performance drops significantly under the black-box setting (i.e., when the attacker cannot access the feature extractor), while \this maintains its capability under all scenarios. Figure~\ref{fig: qvalue} shows the max Q-Value during the training phase. \this achieves $90\%$ convergence in less than $420$ episodes in~\ref{fig: qvalue advanced} and converges within $300$ episodes in~\ref{fig: qvalue classic}, demonstrating strong convergence ability. This ensures that the RL stage incurs low training overhead. We can define the Q-Value threshold $\xi'$ according to the training curve.

\begin{figure}[t]
    \subfigcapskip=-1mm
    \vspace{0mm}
    \begin{center}
    \subfigure[FlowLens / Zeus.]{
        \hspace{-2mm}
        \label{fig: relationship between ASR and threshold, FlowLens}
		\includegraphics[width=0.224\textwidth]{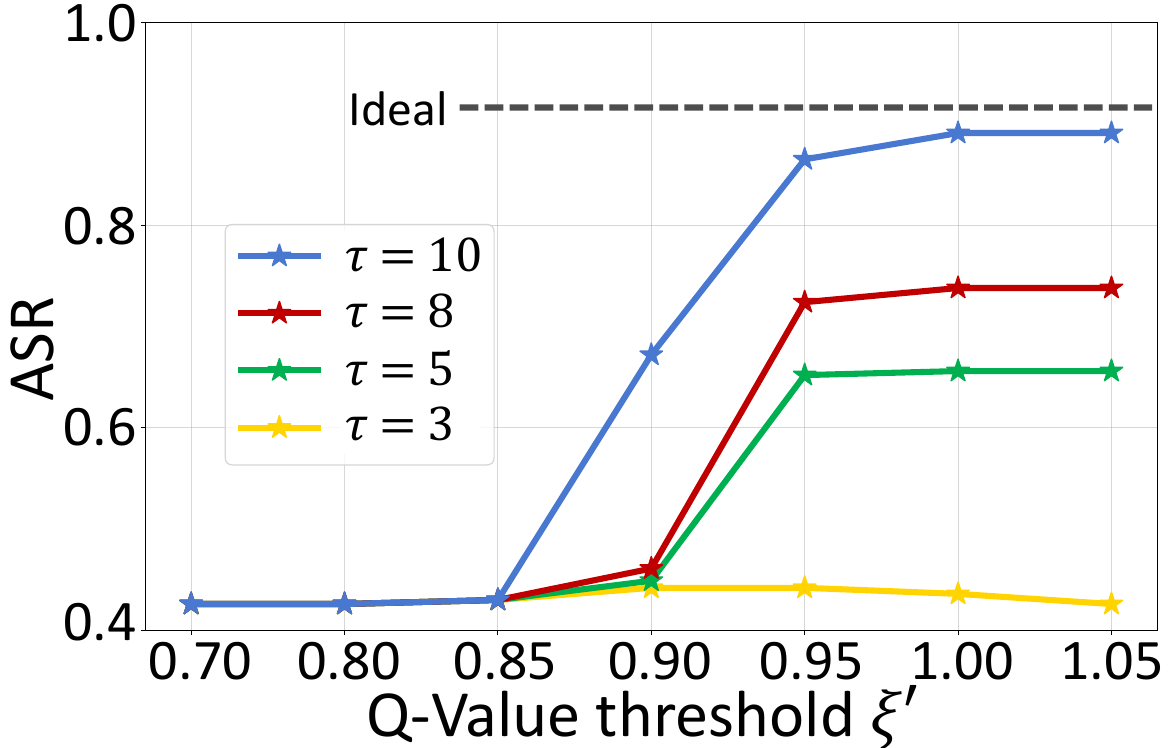}
	}
	\subfigure [Vanilla + RNN / Storm.]{
        \hspace{-2mm}
        \label{fig: relationship between ASR and threshold, RNN}
		\includegraphics[width=0.224\textwidth]{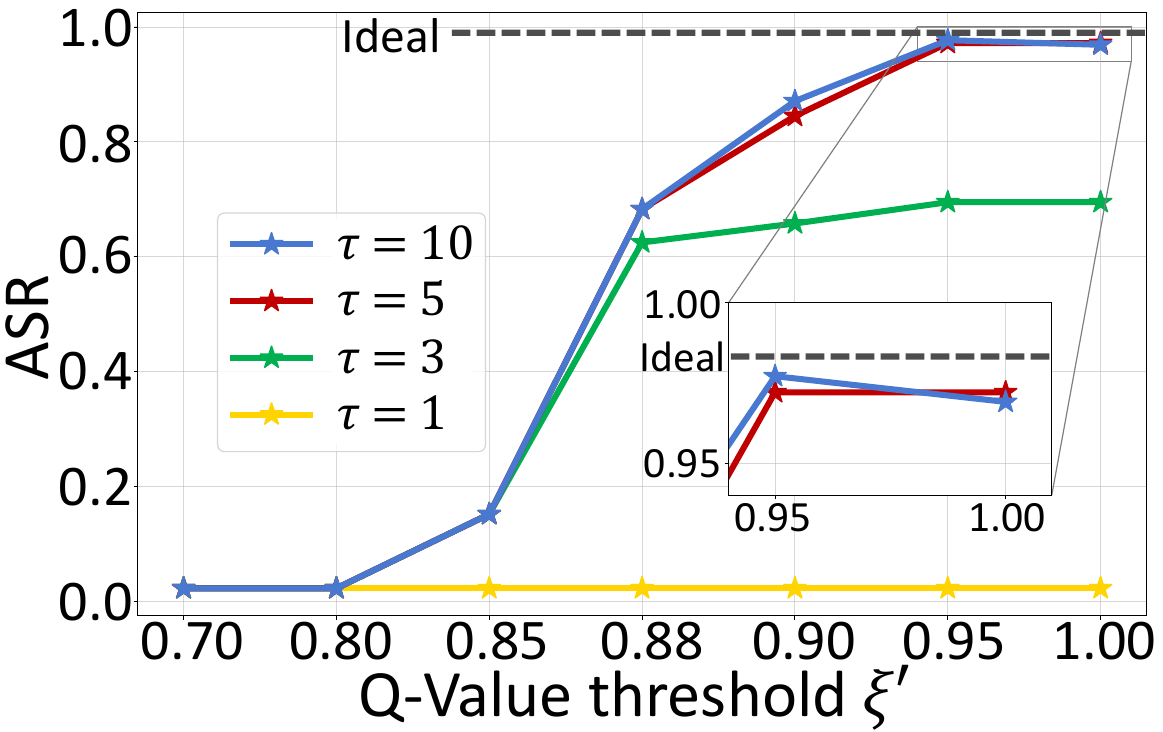}
	}
	\vspace{-2mm}
    \caption{\textbf{The relationship between ASR and Q-Value threshold $\xi'$ under different step thresholds $\tau$.} ``Ideal'' represents the maximum ASR when the attacker can obtain feedback.}
    \label{fig: relationship between ASR and threshold}
    \end{center}
    \vspace{-4mm}
\end{figure}

\this is a stealthy attack capable of generating effective adversarial traffic with minimal modifications. We set the step threshold $\tau$ case-by-case while ensuring it is no more than $10$. Random Mutation and Mutate-and-Inject perform poorly under the same step setting, as shown in Table~\ref{tab: asr comparison}. Amoeba, Traffic Manipulator and other traffic obfuscation methods~\cite{NDSS22-ditto} typically modify most of the packets, making the attack easy to detect. Figure~\ref{fig: relationship between ASR and threshold} illustrates the relationship between ASR and Q-Value threshold $\xi'$ under different step thresholds $\tau$. In Figure~\ref{fig: relationship between ASR and threshold, FlowLens}, we show the effect of attacking FlowLens on the Zeus dataset, which is a complex scenario (i.e., $\text{ASR} = 0.9102$). NetMasquerade achieves near-optimal attack rates within no more than $10$ steps of modifications. According to Figure~\ref{fig: relationship between ASR and threshold, RNN}, we find that ASR is not always positively correlated with the number of modification steps at the same $\xi$, especially in the neighborhood of higher ASR values. An excessively high $\tau$ may lead the algorithm to make excessive modifications when the $\xi$ cannot be satisfied.

\begin{figure}[t]
    \subfigcapskip=-1mm
    \vspace{-4mm}
    \begin{center}
    \subfigure[SSDP Flood.]{
        \hspace{-2mm}
        \label{fig: bandwidth_ssdp}
		\includegraphics[width=0.225\textwidth]{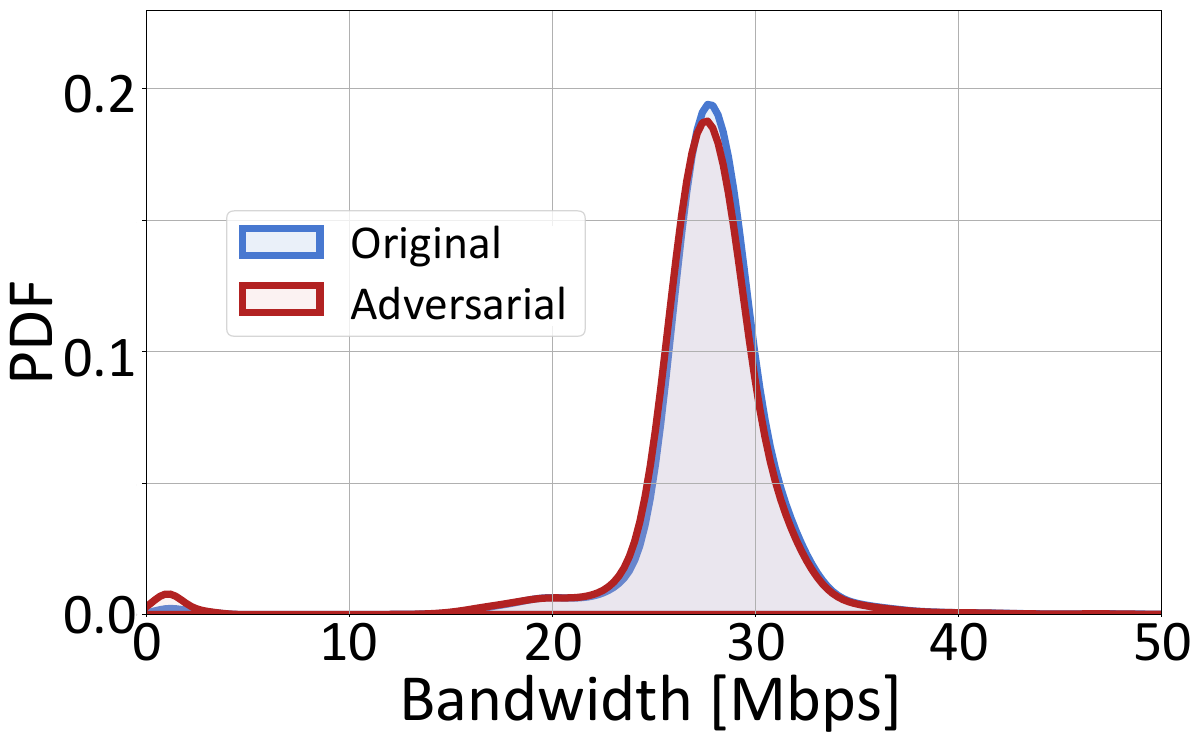}
	}
	\subfigure [SYN DoS.]{
        \hspace{-2mm}
        \label{fig: bandwidth_syn}
		\includegraphics[width=0.225\textwidth]{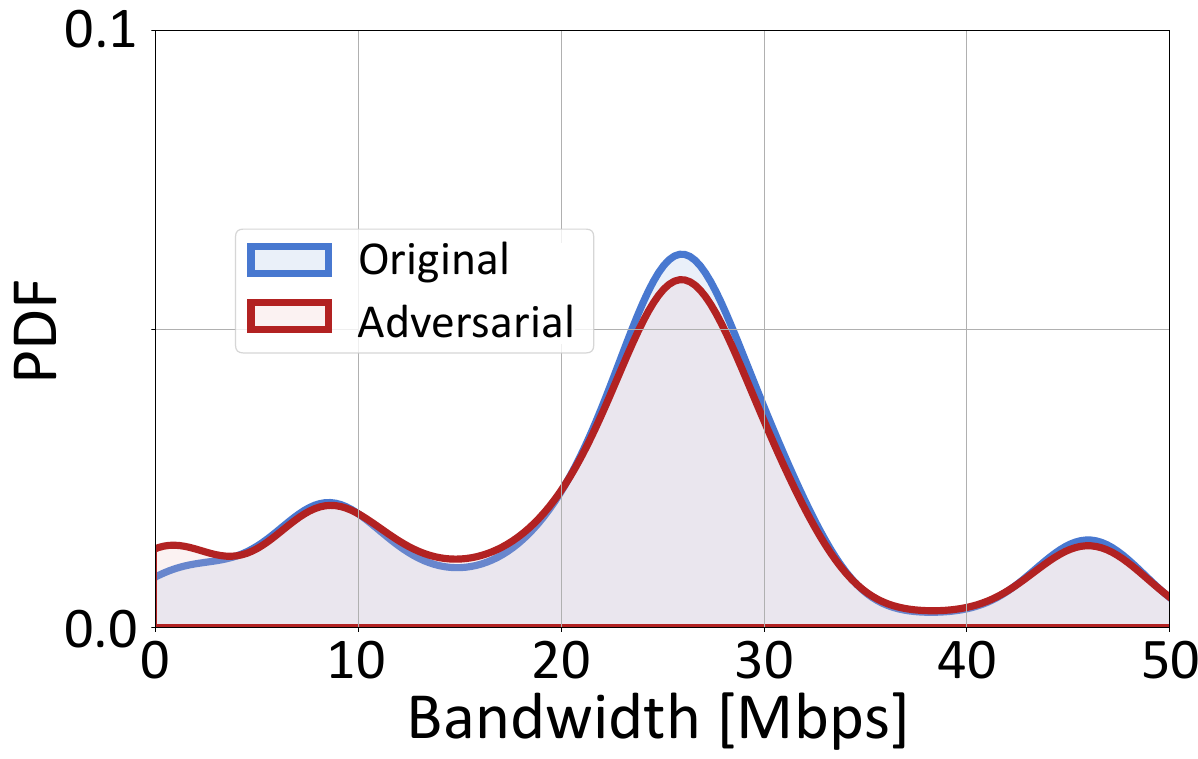}
	}
	\vspace{-2mm}
    \caption{\textbf{Bandwidth of DoS attack.}} 
    \label{fig: bandwidth}
    \end{center}
    \vspace{-6mm}
\end{figure}

\this ensures the effectiveness of adversarial traffic. Our focus is on two types of malicious traffic: SSDP Flood and SYN DoS, whose effectiveness is demonstrated by high rates. We define the effectiveness penalty $r_M$ as the sending rate of the flow. Additionally, by post-processing \model (see \S~\ref{chapter: post-processing} for details), we can limit the range of IPD perturbations. We measure the bandwidth distributions of the original and adversarial traffic, as shown in Figure~\ref{fig: bandwidth}. The KL divergence between the adversarial and original traffic distributions for SSDP Flood and SYN DoS is $0.009$ and $0.013$, indicating that \this does not significantly change the bandwidth distribution. On the other hand, the negligible KL divergence confirms that \this does not introduce any new, delay‑related artifacts that can be readily detected.

\subsection{Overhead and Efficiency}
\noindent \textbf{Training Overhead.} To be practical in real-world scenarios, our attack must be prepared efficiently. Our framework achieves this through the two-stage design that separates computationally intensive training from rapid online execution.

\begin{itemize}[leftmargin=*]
    \item \textbf{Stage 1 (Benign Traffic Pattern Mimicking)} is pre-trained offline on publicly available benign traces. Since this stage can be completed well before the actual attack commences, its end-to-end training cost (about $ \sim 75 $ hours on our testbed) does not affect the online generation speed.
    \item \textbf{Stage 2 (Adversarial Traffic Generation)} runs online during the actual attack setup. We measure the time required for the RL loop (i.e., adversarial flow generation $\rightarrow$ DPDK emission $\rightarrow$ feedback reception $\rightarrow$ policy update) to converge. On our testbed, this stage is highly efficient, with the policy typically converging within just $1$ hour.
\end{itemize}

Overall, the two-stage design shifts most of the overhead to offline training, thereby ensuring timely attack execution.

\begin{figure}[t]
    \subfigcapskip=-3mm
    \vspace{0mm}
    \begin{center}
	\subfigure[Throughput comparison.
    ]{
        \hspace{-3mm}
        \includegraphics[width=0.48\textwidth]{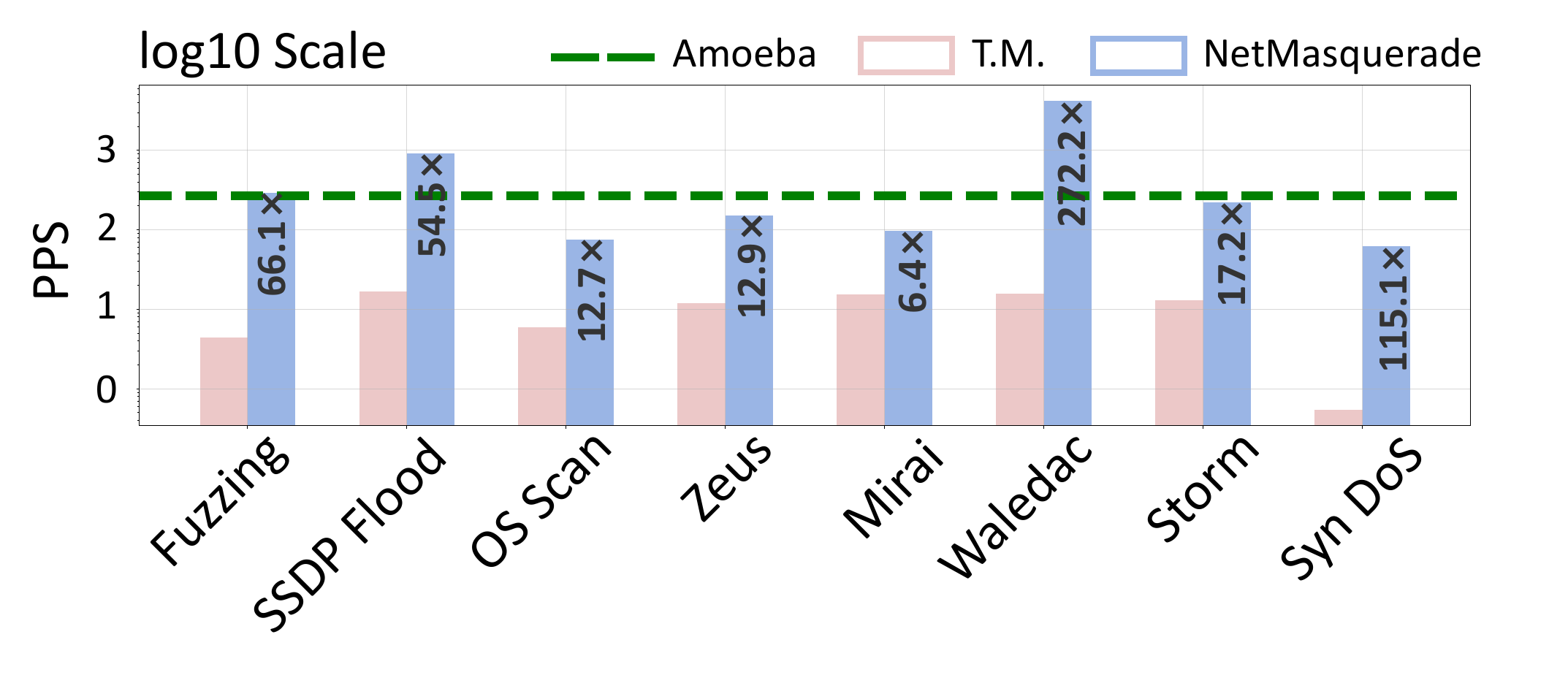}
        \label{fig: Throughput Comparison}
    }
    \\
    \vspace{-3mm}
    \subfigure [Throughput vs. steps under attack 
    ]{
        \hspace{-3mm}
        \includegraphics[width=0.48\textwidth]{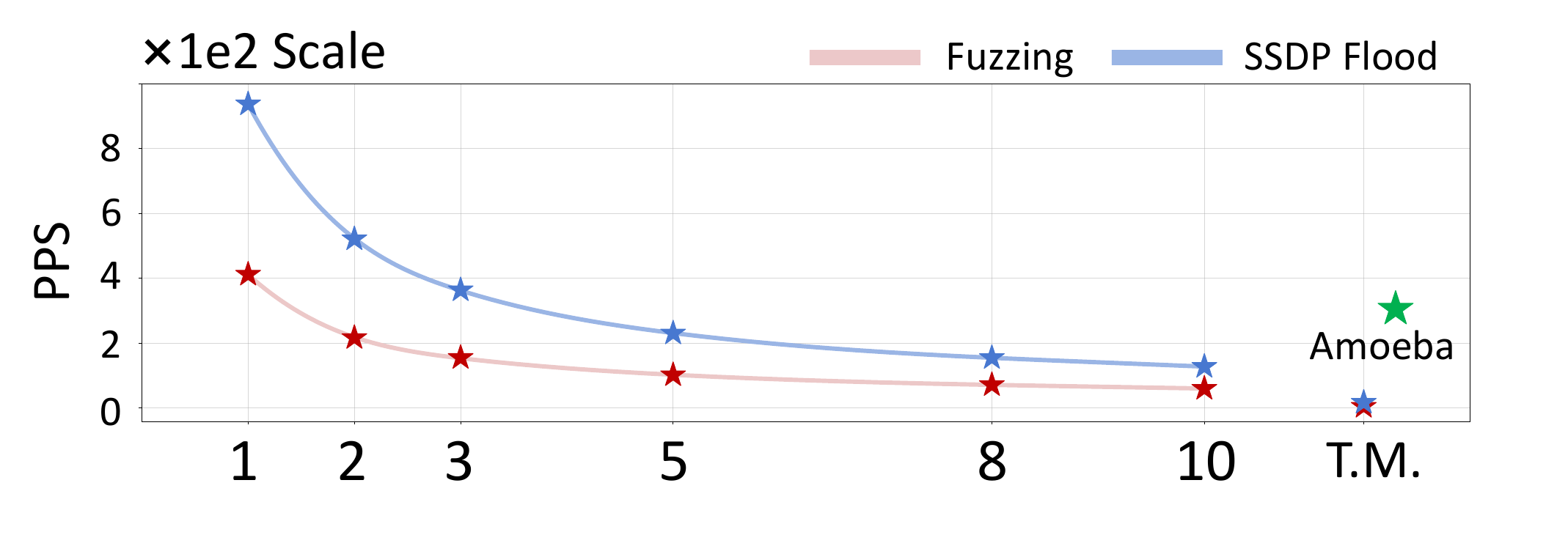}
        \label{fig: Step Throughput}
    }
    \vspace{-4mm}
    \caption{\textbf{Efficiency of \this and baselines.}}
    \label{fig: Throughput}
    \end{center}
    \vspace{-6mm}
\end{figure}

\noindent \textbf{Efficiency.} High inference speed is crucial for generating adversarial traffic, especially in high-throughput network environments. Our measurement is end-to-end, including packet extraction and transmitting packets via our DPDK engine. Figure~\ref{fig: Throughput Comparison} compares the throughput of \this with baseline methods. Both \this and Traffic Manipulator operate at the flow level, yet \this achieves an average efficiency improvement of $\sim 69.6\times$ over Traffic Manipulator across eight datasets. In contrast, Amoeba performs packet-level inference, maintaining a throughput of approximately $300\pm10$ PPS under various attack scenarios. Notably, for long-flow attacks (e.g., Waledac), \this exhibits a clear advantage in efficiency. Figure~\ref{fig: Step Throughput} shows the efficiency curve under different maximum inference steps. By adjusting thresholds $\xi'$ and $\tau$, we can make a trade-off between attack accuracy and inference efficiency. In our experiments, the maximum number of inference steps is generally no more than $10$.

\subsection{Deep Dive} 
\label{deep dive}

\begin{figure}[t]
    \subfigcapskip=-1mm
    \vspace{0mm}
    \begin{center}
    \subfigure[NetBeacon.]{
        \hspace{-2mm}
        \label{fig: ASR under limited probes, NetBeacon}
		\includegraphics[width=0.229\textwidth]{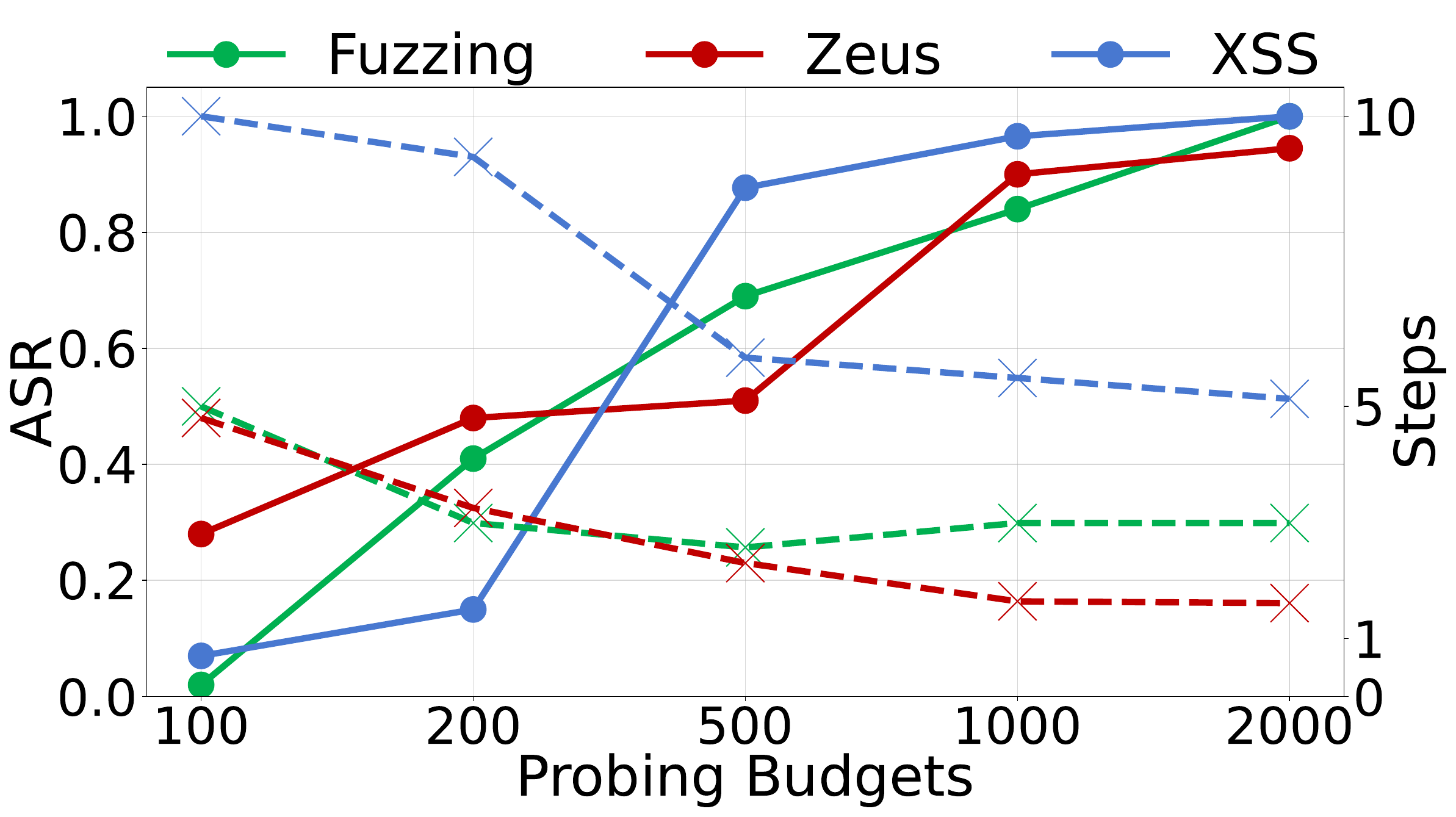}
	}
	\subfigure [Vanilla + RNN.]{
        \hspace{-2mm}
        \label{fig: ASR under limited probes, Vanilla}
		\includegraphics[width=0.229\textwidth]{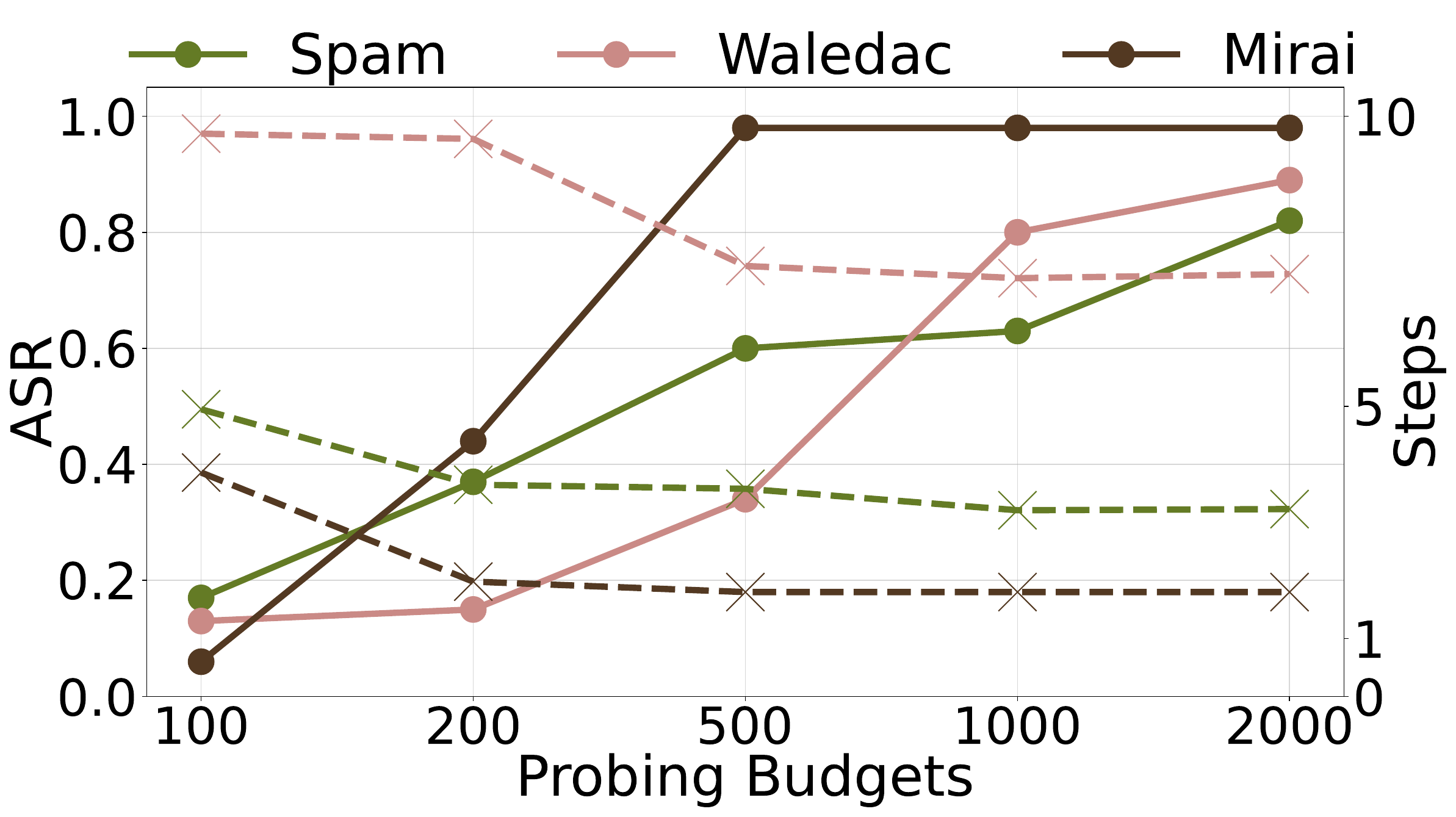}
	}
	\vspace{-4mm}
    \caption{\textbf{ASR under different probing budgets.} Solid lines denote ASR; dashed lines denote steps.}
    \label{fig: ASR vs Probe Budgets.}
    \end{center}
    \vspace{0mm}
\end{figure}

\noindent \textbf{Effectiveness under Limited Probes.} \this assumes that the attacker is able to perform a probing process to collect feedback for model training. However, the probing budget may be limited, as excessive probes could trigger alarms and lead to aggressive countermeasures. To quantify this trade-off, we evaluate the ASR of \this given various probing budgets against two detection systems (NetBeacon and Vanilla + RNN) across three distinct datasets each. As shown in Figure~\ref{fig: ASR vs Probe Budgets.} (the solid lines), the ASR exhibits a steep learning curve between $200$ and $1,000$ probes. For instance, against NetBeacon, NetMasquerade achieves, on average, $35.6\%$, $70.4\%$, and $88.9\%$ of its final ASR with budgets of $200$, $500$, and $1,000$ probes, respectively. The policy typically converges between $1,000$ and $2,000$ probes, although certain complex scenarios (e.g., Vanilla + RNN / Spam) may require a larger budget. This probing load is several orders of magnitude lower than the steady-state workload of a single data-center server (${\sim}500$ flows per second)~\cite{SIGCOMM15-datacenter} or the capacity of a commercial telemetry system~\cite{CiscoTelemetry2024} (over $50,000$ flows per-second~\cite{ciscoSCA}). Crucially, this volume also sits below the thresholds that mainstream IDS products use to raise scan or anomaly alarms~\cite{cisco_CSM_4.19}.

Furthermore, we examine how the probing budget affects the average number of perturbation steps per flow (the dashed lines in Figure~\ref{fig: ASR vs Probe Budgets.}). Since each step costs one probe, this count reflects the model's probe utilization efficiency. In the early training phase, the agent requires more steps to explore the action space and identify effective modifications; as training progresses and feedback accumulates, the average steps per flow drop sharply. This shows the policy is learning to use the budget more efficiently, which is key to learning optimal, minimal-modification policies and contributes to the fast convergence of our framework.

\begin{figure}[t]
    \subfigcapskip=-1mm
    \vspace{-6mm}
    \begin{center}
    \subfigure[Vanilla + RNN.]{
        \hspace{-4mm}
        \label{fig: ASR under noise, NetBeacon}
		\includegraphics[width=0.229\textwidth]{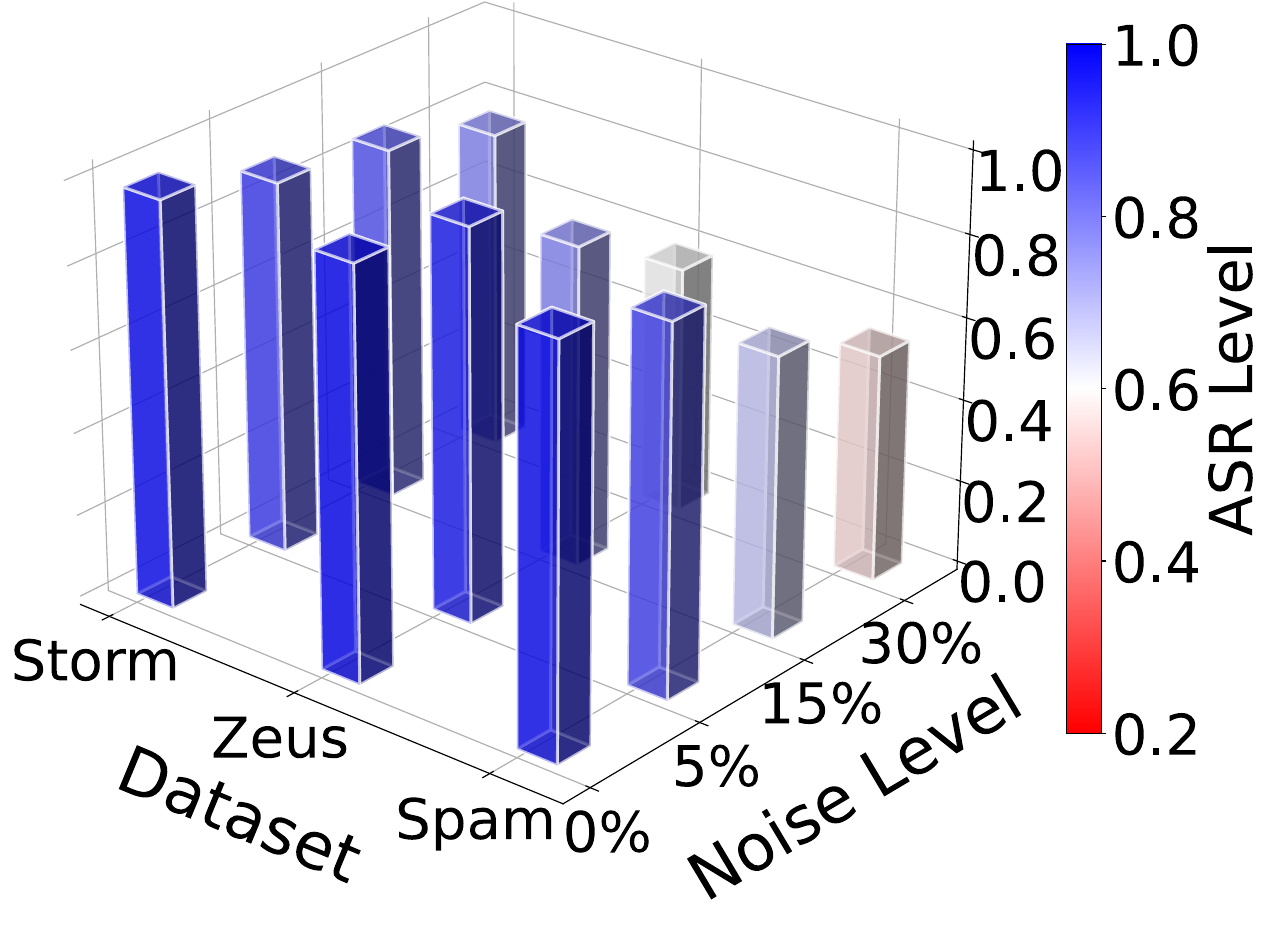}
	}
	\subfigure [Whisper.]{
        \hspace{-2mm}
        \label{fig: ASR under noise, Whisper}
		\includegraphics[width=0.229\textwidth]{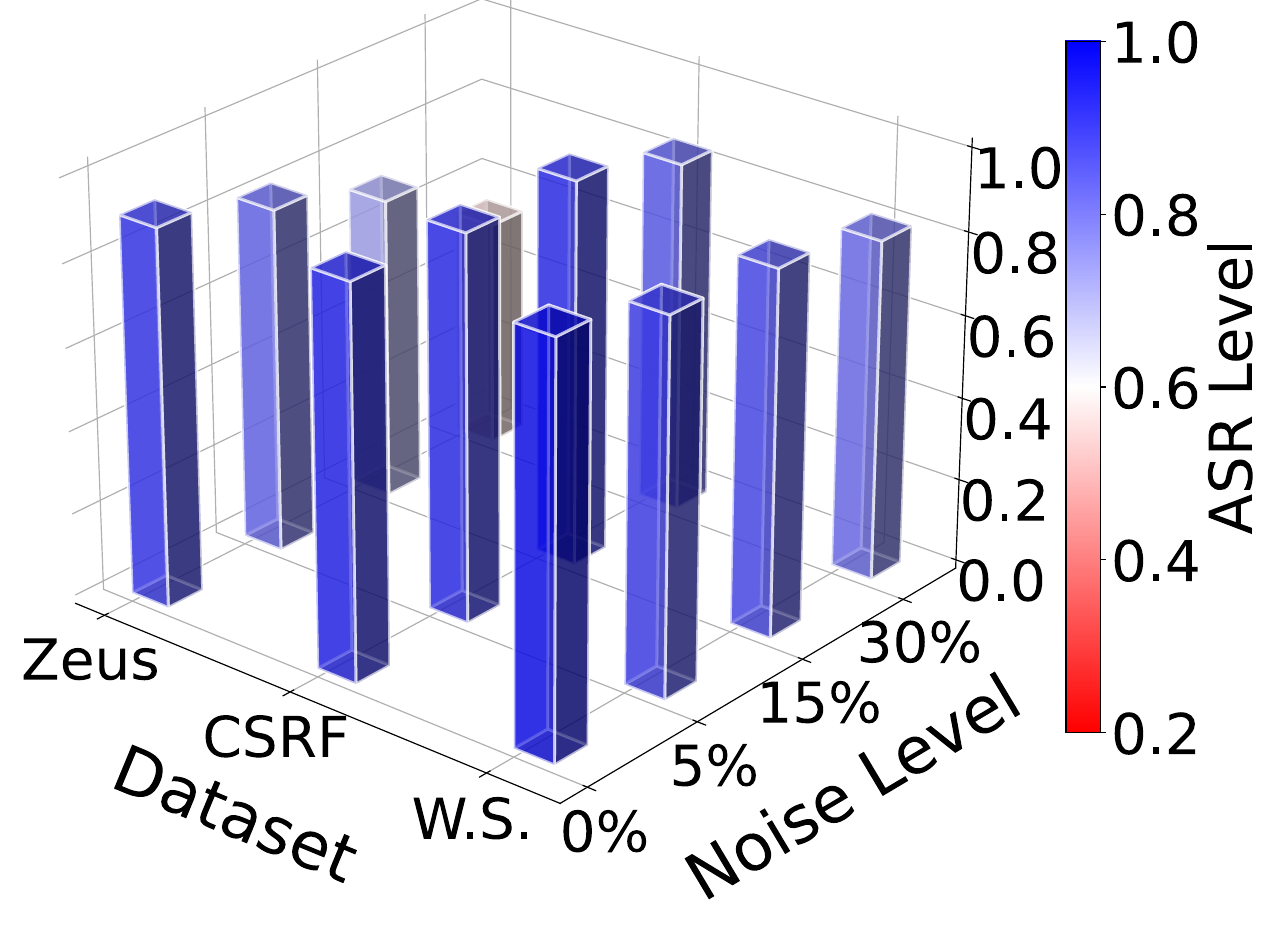}
	}
	\vspace{-2mm}
    \caption{\textbf{ASR under different noise levels.}}
    \label{fig: ASR vs noise.}
    \end{center}
    \vspace{-7mm}
\end{figure}

\noindent \textbf{Robustness to Noisy Feedback.} The feedback that an attacker can collect may be unreliable in real-world deployment. Such noise may be caused by various reasons, including misclassifications, specific configurations, or even adversarial responses from the target designed to mislead the attacker. Figure~\ref{fig: ASR vs noise.} shows the final ASR achieved against the Vanilla + RNN and Whisper under noise levels of $5\%$, $15\%$, and $30\%$. The results demonstrate that NetMasquerade maintains high efficacy under moderate noise. For example, when the noise level is $5\%$, NetMasquerade exhibits an average drop in attack success rate of only $0.063$. When the reward signal becomes highly unreliable (i.e., given a $30\%$ noise level), the ASR degrades gracefully rather than collapsing. Meanwhile, we observe that increased noise primarily affects the convergence efficiency, particularly the stability of the Q-value. With extended explorations, it is still possible to achieve higher ASR.

\noindent \textbf{Importance of \this's Components.} \this's success relies on its two-stage framework design. We validate the necessity of each stage through a series of ablation studies, which show that removing or replacing either stage significantly reduces the attack success rate. Detailed ablation analysis can be found in Appendix~\ref{subsection: Deep Dive Appendix}.

\subsection{Robustness against Defenses}

\begin{figure}[t]
    \subfigcapskip=-2.4mm
    \begin{center}
    \hspace{-2mm}
    \includegraphics[width=0.48\textwidth]{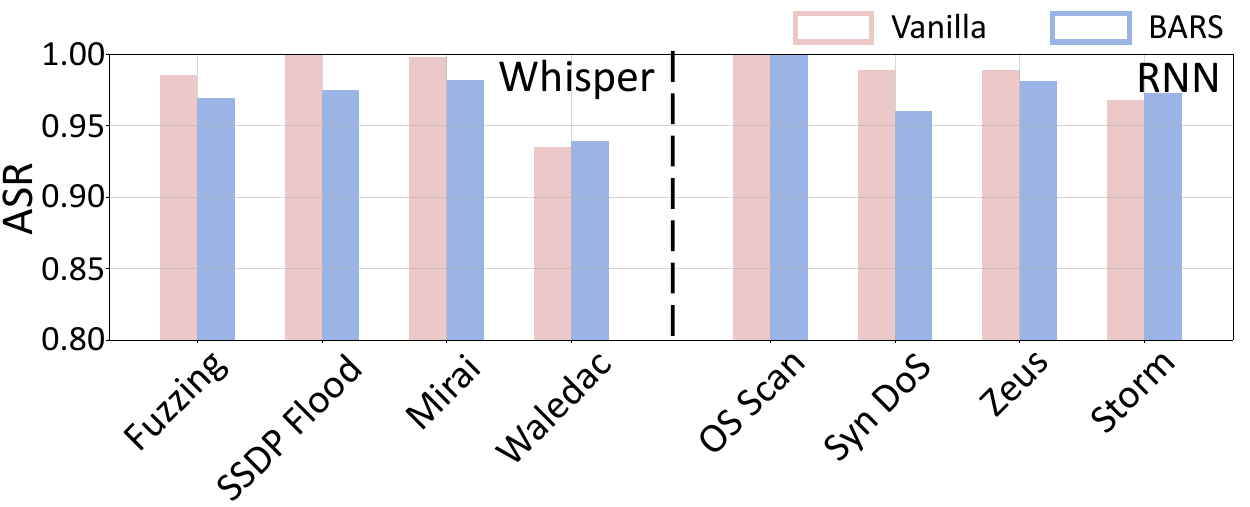}
	\vspace{-3mm}
    \caption{\textbf{ASR against BARS.}}
    \label{fig: BARS}
    \end{center}
    \vspace{-6mm}
\end{figure}

\this generates adversarial traffic within the traffic (physical) space, which translates into an unrestricted adversarial attack in the feature space. Consequently, defenses that operate in the feature space, that is, defenses designed to make a model robust to variations within a certain numerical norm of its input feature vectors, such as adversarial training~\cite{ICLR15-FGSM, S&P16-Distillation} or sample randomized smoothing~\cite{ICML19-RandomizedSmoothing, S&P19-DP}, fail to counter \this.

\begin{itemize}[leftmargin=*]
    \item \textbf{BARS}~\cite{NDSS23-BARS}. BARS employs a combined distribution transformer in feature space to map smoothed noise onto arbitrary manifolds, providing certified robustness bounds for certain perturbations in that space. We use the open-source implementation of BARS and deploy it on two detection methods: Whisper and Vanilla + RNN, and the results are shown in Figure~\ref{fig: BARS}. \this maintains a high ASR across $8$ test datasets even against BARS. 
The primary reason lies in BARS's limited certified bound in the feature space, whereas \this performs multi-step manipulations in the traffic space (e.g., inserting additional packets), leading to perturbations that exceed BARS's bound when projected back into the feature space. Moreover, in some datasets, \this even attains a higher ASR under BARS, likely because random noise reduces the model's accuracy, even with training data augmentation. 
\end{itemize}

\noindent \textbf{Guidelines for Building More Robust Detection Systems}. Overall, it is difficult for the feature‑space defenses to defend against \this, because even small perturbations in traffic space can lead to large distances in feature space. A straightforward countermeasure is traffic‑space adversarial training: augment the training set with packet‑level perturbations to strengthen the model's decision boundaries. Second, instead of certifying an $\mathcal{L}_p$ norm in feature space, traffic space certification (e.g., bounding the number of inserted packets) may be more effective against \this. Finally, \this's training process implicitly assumes a static decision boundary from the target model. Thus, the defense models can introduce randomness at inference to avoid static decision boundaries, such as altering the model architecture~\cite{NIPS22-RandomAgg} or parameters~\cite{NIPS23-RandomWeight}. Such techniques would force the attacker to optimize against a distribution of models rather than a single target, expanding the exploration space. We will explore these strategies in future work.
\section{Related Work}\label{chapter:survey}

\noindent \textbf{ML-based Malicious Traffic Detection.} For generic detection, various methods have been developed to learn flow-level features, such as frequency domain features~\cite{CCS21-Whisper}, distribution features~\cite{NDSS21-FlowLens}, statistical features~\cite{CoNEXT22-Xatu}, and graph features~\cite{NDSS23-HV}.
In particular, existing methods utilize programmable network devices to achieve high efficiency, e.g., NetBeacon~\cite{Security23-NetBeacon} installed decision trees on programmable switches, N3IC implemented binary neural networks on SmartNICs~\cite{NSDI22-N3IC}.
In contrast to these flow-level detections, Kitsune~\cite{NDSS18-Kitsune}, nPrintML~\cite{CCS21-nPrintML}, and CLAP~\cite{Conext20-CLAP} learned packet-level features of malicious traffic.
For task-specific detection, several studies aimed to detect malware behaviors. For example, Tegeler~\etal~\cite{CONEXT12-BotFinder} detected communication traffic from botnets. Similarly, Tang \etal~\cite{INFOCOM20-Zerowall} detected malicious web traffic. Dodia~\etal~\cite{CCS22-TorDetection} identified malicious Tor traffic from malware. Furthermore, Sharma~\etal~\cite{CoNEXT22-Lumen} and Tekiner~\etal~\cite{NDSS22-IoTCryptojacking} captured attack traffic targeting IoT devices.

\noindent \textbf{On the Robustness of Traffic Detection.} Robustness issues are prevalent in traffic analysis systems, i.e., attackers can easily construct adversarial traffic examples to trick the systems into misclassifying traffic. First, Fu~\etal~\cite{TON23-Whisper} revealed that attackers can easily mimic benign traffic to evade the traditional methods~\cite{NDSS18-Kitsune, NSDI22-N3IC} by simply injecting random noise. Such observations necessitate robust detection methods~\cite{NDSS23-HV, CCS21-Whisper, TON23-Whisper}. Second, advanced evasion strategies are developed, which optimize the adversarial traffic examples according to the outputs of white-box~\cite{TDSC22-Manda, CCS21-DomainConstraintsRobustness, IDS20-NAttack} and grey-box detection models~\cite{JSAC21-TM}. These methods are different from our hard-label black-box evasion. Additionally, existing studies analyzed the robustness of traffic analysis other than traffic detection, e.g., improving robustness for web fingerprinting~\cite{Security21-BLANKET, CCS23-Realistic-WF}, which are orthogonal to our evasion attack.

\noindent \textbf{Common Issues of ML-Based Security Application.} 
Sommer~\etal~\cite{SP10-Outside} analyzed why ML-based traffic detection systems suffer from low usability, and emphasized the importance of considering evasion behaviors of real-world attackers. Arp~\etal~\cite{SEC22-Dos-Donots} explored the practical challenges associated with ML-based applications, highlighting issues of evasion attacks~\cite{NDSS18-Kitsune, CCS17-Deeplog}. Moreover, Alahmadi~\etal~\cite{SEC19-99FP}, Vermeer~\etal~\cite{CCS23-Alert-Alchemy}, and Fu~\etal~\cite{CCS23-pVoxel} further demonstrated that existing ML-based traffic detections raised massive false positive alarms. Additionally, Han~\etal~\cite{CCS21-DeepAID}, Jacobs~\etal~\cite{CCS22-Emperor}, and Wei~\etal~\cite{SEC23-xNIDS} addressed the explainability of traffic detection systems.

\section{Conclusion}\label{chapter:conclusion}
In this paper, we introduce \this, a hard-label black-box evasion attack method specifically devised for malicious traffic detection systems. \this employs a two-stage framework.
First, \this establishes a tailored pre-trained model called \model for capturing diverse benign traffic patterns. Subsequently, \this integrates the \model into an RL framework, effectively manipulating malicious packet sequences based on benign traffic patterns with minimal modifications. Also, \this introduces dissimilarity and effectiveness penalties, allowing adversarial traffic to retain attack stealth and effectiveness.
Extensive experiments show that \this enables both high-rate and low-rate attacks to evade $6$ top-performing detection systems in $80$ attack scenarios, achieving over $96.65\%$ attack success rate on average. Moreover, \this applies minimal modifications to no more than $10$ steps in all scenarios. Additionally, \this can achieves
low-latency adversarial traffic generation, demonstrating its practicality in real-world scenarios.
\section{Ethical Considerations}

We carefully assess several ethical aspects to ensure that our study adheres to ethical standards. This work is aimed solely at assessing and improving the robustness of traffic detection models, rather than facilitating malicious or unlawful activities. We strictly follow all terms of use, and no private or sensitive data is accessed or disclosed. All analysis relies exclusively on publicly available datasets—Kitsune, PeerRush, HyperVision (malicious traffic), and MAWI (benign traffic)—without intercepting or manipulating any real-world network traffic. Likewise, we do not perform any active attacks or evasions against real detection systems, ensuring no impact on actual network traffic or stability.

\section*{Acknowledgment}

We thank the anonymous reviewers for their thoughtful comments. This work was supported in part by the National Science Foundation for Distinguished Young Scholars of China (No. 62425201), the National Natural Science Foundation of China (Grant Nos. 62472036, 62472247, 62202258, 62221003, 62132011, 61932016, and U22B2031), and Beijing Nova Program. Yi Zhao and Ke Xu are the corresponding authors.

\bibliographystyle{IEEEtranS}
\bibliography{reference}

\section*{Appendix} \label{chapter:appendix}

\renewcommand{\arraystretch}{1.2}
\begin{table*}[!t]
    \small
    \centering
    \caption{Details of Malicious Traffic Datasets}
    \vspace{0mm}
    \resizebox{\textwidth}{!}{
    \begin{threeparttable}
    \begin{tabularx}{\textwidth}{@{}c|c|>{\centering}X|c|c|c|c|c@{}}
    \toprule
    \multicolumn{2}{c|}{\tabincell{c}{Malicious\\Traffic Dataset}}  & Description  & Source  & Bandwidth & \tabincell{c}{Enc.\\Ratio} &  \tabincell{c}{Mal.\\Ratio} &  \tabincell{c}{External\\Data~\tnote{1}}\\
    \midrule
    \multirow{2}{*}{Recon.}      &  OS Scan                         & Scanning for active hosts and operating systems.     & Kitsune~\cite{NDSS18-Kitsune}    & 0.96 Mbps  & 0.0\%           & 0.0045   & N/A           \\
                                 &  Fuzz Scan                       & Scanning for vulnerabilities in protocols.           & Kitsune                          & 27.9 Mbps  & 0.0\%           & 0.0089   & N/A           \\
    \midrule
    \multirow{2}{*}{DoS}         &  SSDP DoS                        & Amplifying SSDP traffic to flood targets.            & Kitsune                          & 27.2 Mbps  & 0.0\%           & 0.0321   & N/A           \\
                                 &  SYN DoS                         & Flooding servers with half‐open TCP connections.     & Kitsune                          & 23.5 Mbps  & 0.0\%           & 0.0858   & N/A           \\
    \midrule
    \multirow{4}{*}{Botnet}      &  Mirai                           & Infects IoT devices with the Mirai malware.          & Kitsune                          & 0.12 Mbps  & 0.0\%           & 0.8408   & MAWI~\tnote{2}  \\
                                 &  Zeus                            & Botnet infected by a Zeus trojan.                  & PeerRush~\cite{DIMVA13-PeerRush} & 0.06 Mbps  & 0.0\%           & 0.9999   & MAWI          \\
                                 &  Storm                           & Peer‐to‐peer botnet spreading malware.               & PeerRush                         & 25.3 Mbps  & 0.0\%           & 0.9628   & MAWI          \\
                                 &  Waledac                         & Spam botnet harvesting user data.                    & PeerRush                         & 13.9 Mbps  & 0.0\%           & 1.0000   & MAWI          \\
    \midrule  
    \multirow{4}{*}{\tabincell{c}{Enc.\\Web\\Attacks}}  &  Webshell  & Malicious script enabling remote server control.     & H.V.~\cite{NDSS23-HV}     & 11.2 Mbps    & 100.0\%         & 0.0234   & N/A         \\
                                 &  XSS                             & Injects malicious scripts into legitimate websites.  & H.V.                      & 31.8 Mbps    & 100.0\%         & 0.0259   & N/A         \\
                                 &  CSRF                            & Fools authenticated users into unintended actions.  & H.V.                      & 7.73 Mbps    & 100.0\%         & 0.0236   & N/A         \\
                                 &  Spam                            & Bulk messages with phishing / malware.  & H.V.                      & 36.2 Mbps    & 100.0\%         & 0.0238   & N/A         \\

    \bottomrule   
    \end{tabularx}
    \begin{tablenotes}
        \scriptsize
        \item[1] We use an external benign dataset when the malicious dataset is nearly $100\%$ malicious.
        \item[2] To ensure a strict black-box setting, we employ the real-world backbone network traces from the WIDE MAWI project’s August 2023 dataset~\cite{WIDE2023} to train the additional models, keeping them entirely separate from the June 2023 traces used to train Traffic-BERT.
    \end{tablenotes}
    \end{threeparttable}
    }
    \label{tab: dataset details}
    \vspace{-4mm}
\end{table*}

\subsection{Details of Datasets}
\label{subsection: details of datasets}
\noindent To closely mirror real-world scenarios, we replay $4$ categories of malicious traffic, totaling $12$ attack types: Reconnaissance \& Scanning, Denial of Service (DoS), Botnet, and Encrypted Web attacks. The details are shown in Table \ref{tab: dataset details}.

\begin{itemize}[leftmargin=*]
    \item \textit{Reconnaissance and Scanning Attacks}. These attacks identify open ports and services across a wide range of servers by sending specific packets, e.g., by sending ICMP echo requests to determine if a host is active. Adversarial traffic made by \this does not influence the effectiveness. Moreover, since these scanning attacks typically do not involve the transmission of payloads, they can bypass payload-based detection methods. Employing a pattern-based detection system is a common way of detecting such attacks. We select two distinct scanning attacks from Kitsune~\cite{NDSS18-Kitsune}: OS Scan and Fuzz Scan.
    \item \textit{Denial of Service Attacks}. DoS attacks incapacitate targeted services by inundating them with an overwhelming volume of requests, depleting resources and rendering the services unavailable. We selected SSDP DoS and SYN DoS traffic from Kitsune. Similar to scanning attacks, payload-based detection methods fall short against DoS attacks; however, detecting these attacks becomes feasible when focusing on the characteristics of their patterns.
    \item \textit{Botnet Attacks}. Botnets, which are large networks of compromised machines, are controlled by attackers via command and control (C\&C) channels to conduct malicious activities~\cite{ACSAC12-Disclose,CONEXT12-BotFinder}. We employ the Mirai dataset from Kitsune, and analyze data from three typical botnets: Zeus, Storm, and Waledac, which were collected by PeerRush~\cite{DIMVA13-PeerRush}.
    \item \textit{Encrypted Web Attacks.} Typically, malicious web traffic is encrypted with HTTPS, concealing its malicious behavior within the packet payload. This encryption prevents traditional rule-based methods unable to detect the traffic. Meanwhile, most traditional ML-based detection systems cannot effectively detect the traffic due to their low packet rates~\cite{NDSS23-HV}. We obtained four common types of Web attack traffic from~\cite{CCS23-pVoxel}, including automated vulnerability discovery (XSS, CSRF), Webshell, and Spam traffic.
\end{itemize}

We replay traffic from diverse sources—covering high and low throughput flows, as well as encrypted and unencrypted streams—to demonstrate NetMasquerade’s general applicability across varying protocols and tasks. For the Scanning, DoS and Encrypted Web attacks, each target detection system is trained on the malicious and benign traces from the corresponding private datasets. For the Botnet attack, the original datasets contain virtually no benign traffic, so we supplement benign flows with traces from the WIDE MAWI project (August 2023). This choice does not compromise the black-box setting, as the Traffic-BERT model is trained on data from June 2023, ensuring that there is no correlation between the distributions of the datasets. To achieve class balance, we train the target models with thousands of malicious flows and an equal number of benign flows. For the one‑class detection system Kitsune, we use only benign samples during training.

\renewcommand{\arraystretch}{1.2}
\begin{table*}[t!]
    \small
    \centering
    \vspace{2mm}
    \caption{\textbf{Malicious Traffic Detection Systems' Performance}}
    \vspace{0mm}
    \newcolumntype{C}{>{\centering\arraybackslash}X}
    \begin{tabularx}{\textwidth}{@{}c|c|CC|CC|CC|CC|CC|CC@{}}
    \toprule
    \multicolumn{2}{c|}{\multirow{3}{*}[-0.3ex]{\shortstack{Malicious\\Traffic Dataset}}} &  \multicolumn{6}{c|}{Traditional ML-based Systems}             & \multicolumn{6}{c}{DL-based Systems} \\
    \cline{3-14}
    \multicolumn{2}{c|}{}                                                                 & \multicolumn{2}{c}{Whisper}  & \multicolumn{2}{c}{FlowLens}   & \multicolumn{2}{c|}{NetBeacon}   & \multicolumn{2}{c}{Vanilla}     & \multicolumn{2}{c}{CIC.}      & \multicolumn{2}{c}{Kitsune}   \\
    \cline{3-14}
    \multicolumn{2}{c|}{}                                                                 & AUC     & F1      & AUC     & F1      & AUC     & F1      & AUC     & F1     & AUC     & F1     & AUC     &  F1    \\
    \midrule
    \multirow{2}{*}{Recon.} & OS Scan                                                     & 0.9978  & 0.9979  & 0.9946  & 0.9947  & 0.9897  & 0.9899  & 0.9720  & 0.9728 & 0.9980  & 0.9980 & 0.9211  & 0.9780 \\ 
                            & Fuzz Scan                                                   & 0.9905  & 0.9899  & 0.9972  & 0.9933  & 0.9913  & 0.9910  & 0.9662  & 0.9663 & 0.9900  & 0.9901 & 0.9952  & 0.9974 \\
    \cmidrule{1-14}
    \multirow{2}{*}{DoS}    & SSDP DoS                                                    & 0.9167  & 0.9231  & 0.9982  & 0.9981  & 0.9995  & 0.9994  & 0.9790  & 0.9790 & 0.9999  & 0.9999 & 0.9900  & 0.9996 \\
                            & SYN DoS                                                     & 0.9879  & 0.9823  & 0.9815  & 0.9800  & 0.9903  & 0.9833  & 0.9616  & 0.9560 & 0.9852  & 0.9849 & 0.9801  & 0.9213 \\
    \midrule
    \multirow{4}{*}{Botnet} & Mirai                                                       & 0.9449  & 0.9458  & 0.9600  & 0.9463  & 0.9444  & 0.9371  & 0.9099  & 0.9156 & 0.9574  & 0.9521 & 0.9322  & 0.9762 \\
                            & Zeus                                                        & 0.9121  & 0.9056  & 0.9250  & 0.8837  & 0.9279  & 0.9516  & 0.9118  & 0.9041 & 0.9625  & 0.9484 & 0.9246  & 0.9017 \\
                            & Storm                                                       & 0.9495  & 0.9468  & 0.9395  & 0.9415  & 0.9972  & 0.9978  & 0.9233  & 0.9271 & 0.9968  & 0.9982 & 0.9302  & 0.9822 \\
                            & Waledac                                                     & 0.9505  & 0.9484  & 0.9660  & 0.9653  & 0.9285  & 0.9467  & 0.9299  & 0.9304 & 0.9860  & 0.9862 & 0.8964  & 0.8414 \\

    \midrule
    \multirow{4}{*}{Enc.}   & Webshell                                                    & 0.9980  & 0.9979  & 0.9955  & 0.9946  & 0.9943  & 0.9955  & 0.9989  & 0.9874 & 0.9965  & 0.9964 & 0.9996  & 0.9887 \\
                            & XSS                                                         & 0.9975  & 0.9974  & 0.9965  & 0.9965  & 0.9967  & 0.9966  & 0.9845  & 0.9937 & 0.9937  & 0.9984 & 0.9991  & 0.9990 \\
                            & CSRF                                                        & 0.9944  & 0.9950  & 0.9920  & 0.9920  & 0.9935  & 0.9934  & 0.9819  & 0.9822 & 0.9928  & 0.9927 & 0.9019  & 0.6625 \\
                            & Spam                                                        & 0.9200  & 0.9135  & 0.9780  & 0.9756  & 0.9635  & 0.9622  & 0.8897  & 0.8847 & 0.9900  & 0.9901 & 0.8887  & 0.8690 \\
    
    \bottomrule
    \end{tabularx}
    \label{tab: acc_compare}
    \vspace{-4mm}
\end{table*}

\subsection{Details of Target Detection Systems}
\label{subsection: details of target systems}
\noindent \textbf{Target Systems.} We use three advanced traditional ML-based malicious traffic detection systems as target systems:
\begin{itemize}[leftmargin=*]
    \item Whisper~\cite{CCS21-Whisper}. Whisper transforms the patterns of flows into frequency domain features, employing clustering to unsupervisedly learn these features. Similarly, the effectiveness of original traffic is assessed by the distance between frequency domain features and the cluster centers. Whisper is particularly effective at detecting DoS traffic, so we retrain the model on the DoS dataset using its default configuration. For botnet traffic, we replace clustering with a linear classifier to enhance detection capabilities.
    \item FlowLens~\cite{NDSS21-FlowLens}. FlowLens samples the distribution of packet-level features on the data plane and uses random forests to learn these features in a supervised manner on the control plane, introducing a new paradigm of malicious traffic detection. We retrain the model with the default model structure and hyperparameters described in the paper.
    \item NetBeacon~\cite{Security23-NetBeacon}. NetBeacon introduces the concept of Intelligent Data Plane (IDP), performing flow-level feature extraction and feature classification with tree-based models directly in the data plane. We reconstruct the feature extraction method described in the paper, select XGBoost~\cite{SIGKDD16-Xgboost} as the representative tree model for traffic classification, and adjust hyperparameters to achieve optimal accuracy.
\end{itemize}

We also implement three top-performing DL-based systems:
\begin{itemize}[leftmargin=*]
    \item CICFlowMeter~\cite{ICISSP16-Gil,ICISSP17-Lashkari} + MLP. CICFlowMeter is a widely used feature processor that extracts over $80$ time-related features from flow patterns. We employ a four-layer linear neural network to learn these features, with the number of neurons in the hidden layers set to three times that of the input layer. By adjusting the hyperparameters, we achieve the model's best classification performance.
    \item Vanilla + RNN. The native feature extractor extracts sequences of sizes and IPDs from the flow, without additional feature processing. Given the sequential nature of the features, we use a single-layer LSTM as the model, taking the concatenation of the two feature sequences as input.
    \item Kitsune~\cite{NDSS18-Kitsune}. Kitsune dynamically extracts and maintains per-packet features through specially designed feature extractors and uses autoencoders and clustering to learn the features of benign traffic unsupervisedly. The detection of malicious traffic is based on the discrepancy between the autoencoder's output and the original feature input. We retrain the model using its original feature extraction and model structure with default hyperparameters.
\end{itemize}

\noindent \textbf{Target System Detection Performance.} Table \ref{tab: acc_compare} summarizes the detection performance of $6$ target systems on $12$ types of malicious traffic. Notably, Kitsune outputs a score indicating how malicious each sample is; we perform a grid search to determine a threshold and compute the corresponding F1 score. We then apply this threshold in the attack experiments to calculate the ASR. All detection systems achieve high AUC and F1 scores, demonstrating strong performance in the absence of evasion attacks.

\renewcommand{\arraystretch}{1.05}
\begin{table}[htbp]
    \centering
    \caption{Details of Hyperparameters.}
    \label{tab: parameters}
    \scriptsize
    \resizebox{\columnwidth}{!}{
    \begin{tabular}{c|c|c|c}
        \toprule
        \textbf{Stage}                                            & \textbf{Hyperparameter}   & \textbf{Value}    & \textbf{Description} \\
        \midrule
        \multirow{7}{*}{\tabincell{c}{\textbf{1}: Traffic-BERT}}  & $n$                       & $512$             & Fixed length of sequences. \\
                                                                  & $d_k$                     & $128$             & Embedding size / total length of $Q, K, V$. \\
                                                                  & \textsc{n\_layers}        & $6$               & Number of encoder blocks. \\
                                                                  & \textsc{attn\_heads}      & $8$               & Number of attention heads. \\
                                                                  & \textsc{d\_ff}            & $512$             & Dimension of feed-forward layer. \\
                                                                  & \textsc{t\_size}          & $56$              & Size of IPD feature vocabulary. \\
                                                                  & \textsc{s\_size}          & $1606$            & Size of size feature vocabulary. \\
        \midrule
        \multirow{8}{*}{\tabincell{c}{\textbf{2}: RL process}}    & $\beta$                   & $0.01 \sim 0.1$   & Weight of $r_D$. \\
                                                                  & $\gamma$                  & $0 \sim 0.2$      & Weight of $r_M$. \\
                                                                  & $\tau$                    & $\leq 10$         & Max step threshold. \\
                                                                  & $\xi'$                    & $0.8 \sim 1.05$   & Stop reward threshold. \\
                                                                  & $\eta$                    & $1$               & Discount factor. \\
                                                                  & $\lambda$                 & $0.9$             & Soft update weight of Q-networks. \\
                                                                  & $\mathcal{B}$             & $1e5$             & Size of experience replay buffer. \\
                                                                  & \textsc{target\_entropy}  & $-10$             & Desired policy entropy (related to $\alpha$) \\
        \bottomrule
    \end{tabular}
    }
    \vspace{-2mm}
\end{table}

\subsection{Details of Hyperparameter Settings}
The default hyperparameters of \this are listed in Table~\ref{tab: parameters}.

\subsection{Deep Dive into \this}
\label{subsection: Deep Dive Appendix}
\noindent \textbf{Importance of Two-Stage Framework.} \this consists of two stages: benign traffic pattern mimicking and adversarial traffic generation. To study the importance of each stage, we design three ablation strategies and conduct attacks on two detection systems, NetBeacon and CICFlowMeter + MLP, across all eight datasets. Table~\ref{tab: Two-Stage Structure effect} shows the ASR.

In the first scenario, we retain stage 1 and replace stage 2 with randomly selecting positions for feature modifications(denoted as $S_1$). Clearly, under this setting, the attacker cannot find the optimal modification positions, resulting in a significant drop in attack capability. On both datasets, the attack success rate drops by an average of $56\%$. This result underscores that merely integrating BERT-generated traffic patterns is insufficient to evade detection; the RL step in Stage 2 is crucial for identifying the most strategic insertion points.

In the second scenario, we remove stage $2$ and replace it with a new Mask-Fill strategy. The first strategy is to fill in the selected positions with stochastic values between the minimum and maximum values of the same flow feature sequence (denoted as $S_2\text{-S}$). By converting the Markov decision process from deterministic to stochastic, it becomes exceedingly difficult for the RL algorithm to converge. Consequently, we observe that the RL strategy predominantly adds chaff packets because the rewards for this type of action are relatively stable.
The second strategy is filling in the selected positions with the average value of the same flow feature (denoted as $S_2\text{-F}$). Due to the variation in sending patterns across different flow segments, this strategy is limited in some cases. As Table 3 illustrates, the attack success rate for these two strategies decrease by an average of $37.4\%$ and $26.8\%$, and neither strategy is effective against high-speed traffic. This outcome underscores how merely relying on an average-value approach or a random approach to features cannot capture dynamic and peak-driven traffic patterns—an issue that becomes even more pronounced in fine-tuning scenarios such as high-speed traffic. Traffic-BERT, on the other hand, guides the RL training process by offering stable and effective benign-pattern perturbations. Although more complex candidate Mask-Fill rules could be considered, these rules can only be applied during stage 2, which would exponentially expand the action space and lead to an action space explosion.

\renewcommand{\arraystretch}{1.2}
\begin{table}[t]
    \footnotesize
    \centering
    \setlength\tabcolsep{2.5pt}
    \caption{\textbf{Effect of Two-stage Framework.} $S_1$, $S_2\text{-S}$, $S_2\text{-F}$, NetM. stand for stage $1$ only, stage $2$ only with stochastic value, stage $2$ only with fixed value, and the overall \this.}
    \vspace{0mm}
    \resizebox{\linewidth}{!}{
    \begin{tabular}{@{}c|cccc|cccc@{}}
    \toprule
    Target Systems & \multicolumn{4}{c|}{NetBeacon} & \multicolumn{4}{c}{CICFlowMeter + MLP} \\
    \midrule
    Methods & $S_1$ & $S_2\text{-S}$ & $S_2\text{-F}$ & NetM.& $S_1$ & $S_2\text{-S}$ & $S_2\text{-F}$ & NetM.\\
    \midrule
    OS Scan & 0.940 & 0.990 & 0.990 & 0.999 & 0.479 & 0.990 & 0.999 & 0.999 \\
    Fuzzing & 0.538 & 0.936 & 0.996 & 0.999 & 0.063 & 0.004 & 0.009 & 0.974 \\
    SSDP Flood & 0.001 & 0.434 & 0.481 & 0.999 & 0.488 & 0.557 & 0.639 & 0.999 \\
    SYN DoS & 0.058 & 0.325 & 0.545 & 0.999 & 0.508 & 0.011 & 0.754 & 0.996 \\
    Mirai & 0.915 & 0.895 & 0.915 & 0.990 & 0.488 & 0.856 & 0.938 & 0.999 \\
    Zeus & 0.355 & 0.508 & 0.508 & 0.945 & 0.347 & 0.813 & 0.891 & 0.987 \\
    Storm & 0.201 & 0.233 & 0.239 & 0.997 & 0.647 & 0.797 & 0.817 & 0.890 \\
    Waledac & 0.750 & 0.727 & 0.924 & 0.999 & 0.123 & 0.783 & 0.880 & 0.981 \\
    \bottomrule
    \end{tabular}}
    \label{tab: Two-Stage Structure effect}
    \vspace{-3mm}
\end{table}

\end{document}